\documentclass[10pt]{iopart}

\usepackage{etoolbox}
\usepackage{hyperref}

\expandafter\let\csname equation*\endcsname=\relax
\expandafter\let\csname endequation*\endcsname=\relax
\usepackage{color}
\usepackage{amsmath}
\usepackage{amssymb}
\usepackage{enumerate}
\usepackage{graphicx}
\usepackage{mathbbol}
\usepackage{amsfonts}
\usepackage{url}

\newcommand{\sgn}{\operatorname{sgn}}

\newcommand{\bea}{\begin{eqnarray}}
\newcommand{\eea}{\end{eqnarray}}
\newcommand{\beq}{\begin{equation}}
\newcommand{\eeq}{\end{equation}}

\def\XXint#1#2#3{{\setbox0=\hbox{$#1{#2#3}{\int}$}
 \vcenter{\hbox{$#2#3$}}\kern-.5\wd0}}

\usepackage[utf8]{inputenc}
\usepackage{amsmath}
\usepackage{hyperref}
\usepackage{graphicx}
\usepackage{amsfonts}
\usepackage{amsthm}
\usepackage{cases}
\usepackage{bm}
\usepackage{amssymb}

\usepackage[a4paper]{geometry}            
\usepackage[english]{babel}                
\usepackage{hyperref}                    

\usepackage{amsmath}    
\usepackage{amssymb}    
\usepackage{amsthm}        
\usepackage{mathtools}

\usepackage{bm}            

\usepackage{color}
\usepackage[table]{xcolor}

\usepackage{graphicx}        
\usepackage{caption}        
\usepackage{subcaption}    

\usepackage{color}
\definecolor{Blue}{rgb}{0.00, 0.00, 1.00}
\definecolor{Red}{rgb}{1.00, 0.00, 0.00}

\hypersetup{
    colorlinks=true,       
    linkcolor=red,          
    citecolor=blue,        
    filecolor=magenta,      
    urlcolor=cyan           
}

\newcommand{\be}{\begin{equation}}
\newcommand{\ee}{\end{equation}}

\newcommand{\beqn}{\begin{eqnarray}}
\newcommand{\eeqn}{\end{eqnarray}}

\makeatletter
\def\@mkboth#1#2{}
\newlength\appendixwidth
\preto\appendix{\addtocontents{toc}{\protect\patchl@section}}
\newcommand{\patchl@section}{%
  \settowidth{\appendixwidth}{\textbf{Appendix }}%
  \addtolength{\appendixwidth}{1.5em}%
  \patchcmd{\l@section}{1.5em}{\appendixwidth}{}{\ddt}%
}
\makeatother

\begin{document}
\title[]{Universal gap statistics for random walks for a class of  
jump densities}

\author{Matteo Battilana}
\address{LPTMS, CNRS, Univ. Paris-Sud, Universit\'e Paris-Saclay, 91405 Orsay, France}

\author{Satya N. Majumdar}
\address{LPTMS, CNRS, Univ. Paris-Sud, Universit\'e Paris-Saclay, 91405 Orsay, France}

\author{Gr\'egory Schehr}
\address{LPTMS, CNRS, Univ. Paris-Sud, Universit\'e Paris-Saclay, 91405 Orsay, France}

\begin{abstract}
We study the order statistics of a random walk (RW) of $n$ steps whose jumps are distributed according to 
symmetric Erlang densities $f_p(\eta)\sim |\eta|^p \,e^{-|\eta|}$, parametrized by a non-negative integer $p$. Our main focus is on the
statistics of the gaps $d_{k,n}$ between two successive maxima $d_{k,n}=M_{k,n}-M_{k+1,n}$ where $M_{k,n}$ is
the $k$-th maximum of the RW between step 1 and step $n$. In the limit of large $n$, we show that the probability density function of the gaps $P_{k,n}(\Delta) = \Pr(d_{k,n} = \Delta)$ reaches a stationary density $P_{k,n}(\Delta) \to p_k(\Delta)$. For large $k$, we demonstrate that the typical fluctuations of the gap, for $d_{k,n}= O(1/\sqrt{k})$ (and $n \to \infty$), are described by a non-trivial scaling function that is independent of $k$ and of the jump probability density function  $f_p(\eta)$, thus corroborating our conjecture about the universality of the regime of typical fluctuations (see G. Schehr, S. N. Majumdar, Phys. Rev. Lett.  {\bf 108}, 040601 (2012)). We also investigate the large fluctuations of the gap, for $d_{k,n}  = O(1)$ (and $n \to \infty$), and show that these two regimes of typical and large fluctuations of the gaps match smoothly. 
\end{abstract}

\maketitle

\tableofcontents

\newpage

%
%
%
%

\section{Introduction}

\begin{figure}[ht]
\centering
\begin{subfigure}{.5\textwidth}
  \centering
  \includegraphics[width=.9\linewidth]{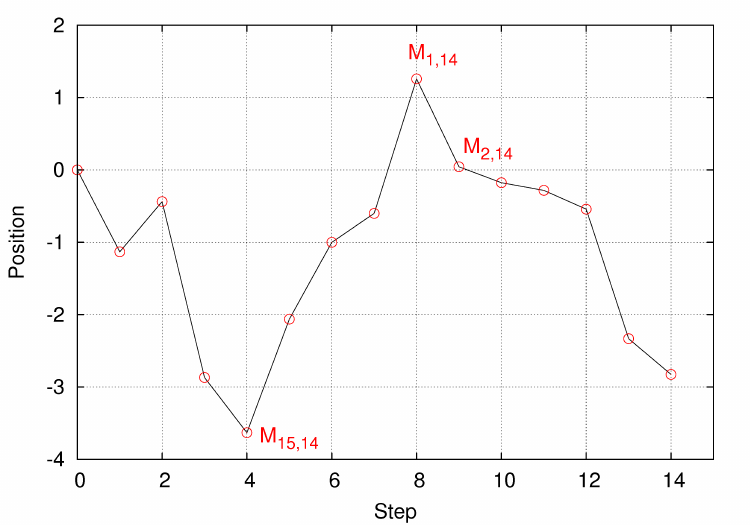}
\end{subfigure}%
\begin{subfigure}{.5\textwidth}
  \centering
  \includegraphics[width=.9\linewidth]{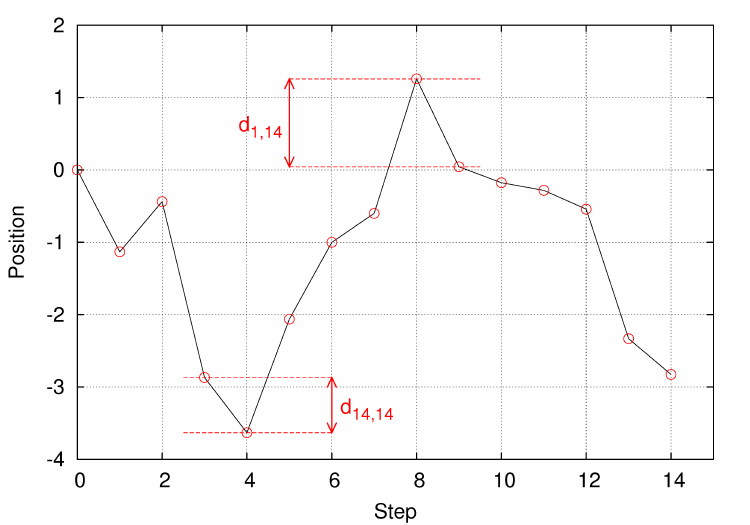}
\end{subfigure}
\caption{{Realization of a RW with $n=14$ steps}. In both we show the the same realization of the RW but in (a) we show the maxima $M_{k,n}$ while in (b) we show the gaps between successive maxima $d_{k,n} = M_{k,n} - M_{k+1,n}$ [see Eq. (\ref{def_gap})].}\label{fig:gaps_maxima}
\end{figure}

Extreme value statistics (EVS) have found many applications in various scientific areas. These include 
environmental science \cite{hydro,fire,eco} or finance \cite{insu,dassios,stock} where extreme events, like tsunamis, earthquakes or financial crashes, although rare, may have disastrous consequences. It has also been realized that EVS plays a major
role in statistical physics, in particular in the context of complex and disordered systems \cite{UC,disord}, random walks and Brownian motions \cite{airy1,RSRG,philippe,review_record} or random matrix theory \cite{TW,top}. Of course, EVS is also a well developed branch of mathematics, at the interface between probability and statistics \cite{gumbel,galambos,LLR2012}. The basic problem of EVS can be formulated as follows. Given a set of $n$ random variables $x_1, x_2, \cdots, x_n$, one is interested in the joint statistics of the $k$-th maxima $M_{k,n}$ among the $x_i$'s, with $M_{1,n} > M_{2,n} > \cdots > M_{n,n}$ (with $M_{1,n} = \max_{1\leq i \leq n} x_i$ and $M_{n,n} = \min_{1\leq i \leq n} x_i$). The statistics of this ordered sequence of $M_{k,n}$'s is usually called {\it order statistics}. Important observables are the gaps between successive maxima~(see Fig. \ref{fig:gaps_maxima})
\beq
\label{def_gap}
d_{k,n}=M_{k,n}-M_{k+1,n} \;,
\eeq
which are useful for instance to characterize crowding effects near the maximum \cite{near_ext1,near_ext2} and have recently been studied in various contexts, notably for branching Brownian motions \cite{derrida1, ramola1,ramola2} as well as for $1/f^\alpha$ signals \cite{racz}, with applications in cosmology \cite{tremaine}.

The theory of order statistics is very well understood in the case where the $x_i$'s are independently and identically distributed (i.i.d.) random variables. These results for i.i.d. random variables are quite robust and actually hold also in the case of weakly correlated random variables, for instance when the correlations between the $x_i$'s decay exponentially \cite{pal}. However, much less is known for {\it strongly} correlated  random variables. A useful and physically relevant example of such strongly correlated time series where extreme value questions can be studied analytically is provided by the positions  
of a random walker (RW), starting at $x_0 = 0$ and evolving according to the Markov rule  
\begin{equation}\label{eq:1}
\begin{split}
x_{i+1}=x_i+\eta_i \;,
\end{split}
\end{equation}
where the jumps $\eta_i$'s are {i.i.d.} random variables drawn from a symmetric and continuous probability density function (PDF) $f_p(\eta)$. The study of order statistics of RWs has a rather long history, since the fifties, in the mathematics literature \cite{FP1952,wendel} as it is an important part of fluctuation theory \cite{Feller}. More recently, order statistics of RWs were revisited by two of us \cite{UOSRW}, with a special focus on the statistics of the gaps (\ref{def_gap}), namely their PDF 
\beq\label{def_pkn}
{\rm Pr}(d_{k,n} \in [\Delta, \Delta + d \Delta]) = P_{k,n}(\Delta)\, d\Delta \;.
\eeq
For the special case of a symmetric exponential density of the jumps $\eta_i$'s 
\beq\label{def_Laplace}
f_0(\eta) = \frac{1}{2b} e^{-|\eta|/b} \;,
\eeq 
an explicit expression for the double generating function (GF) of $P_{k,n}(\Delta)$ with respect to both $n$ and $k$
was obtained [see Eqs. (\ref{exact_A1})-(\ref{GF_pdf_delta}) below]. In particular, in the large $n$ limit, it was shown \cite{UOSRW} that $P_{k,n}(\Delta)$ converges to a limiting density $p_k(\Delta)$, i.e.,
\beq\label{limiting}
P_{k,n}(\Delta) \to p_k(\Delta) \;, \; n \to \infty \;,
\eeq
and an explicit expression for the GF of $p_k(\Delta)$ was obtained [see Eqs. (\ref{stationary}) and (\ref{expl_stationary}) below]. In the large $k$ limit, it was shown that the average size of the gap is $\propto 1/\sqrt{k}$. It turns out that even the fluctuations of the gap are ``typically'' of order $O(1/\sqrt{k})$. This means that the PDF of the gap $\Delta$ has a single scale, $\propto 1/\sqrt{k}$, and takes the scaling form for $\Delta = O(1/\sqrt{k})$
\begin{equation}\label{eq:typ_scaling}
\begin{split}
&p_{k}\left(\Delta\right)\sim \frac{\sqrt{k}}{\sigma_0}P\left(\Delta\frac{\sqrt{k}}{\sigma_0}\right) \;,
\end{split}
\end{equation}
with $\sigma_0^2 = 2 b^2$ being the variance of the jump density in (\ref{def_Laplace}). The scaling function $P(x)$ in Eq. (\ref{eq:typ_scaling}) was computed explicitly as \cite{UOSRW}
\begin{equation}\label{P_scaling}
\begin{split}
P(x) = \left[\sqrt{\frac{2}{\pi}}\left(1+2x^2\right)-e^{2x^2}x\left(4x^2+3\right)\text{erfc}\left(\sqrt{2}x\right)\right] \;,
\end{split}
\end{equation}
where ${\rm erfc}(z) = (2/\sqrt{\pi})\int_z^\infty e^{-t^2} \, dt$ is the complementary error function. In particular, it has the asymptotic behaviors
\beq\label{asympt_P}
P(x) \sim
\begin{cases}
&4\sqrt{2/\pi} \;, \; x \to 0 \\
& \\
&3/\sqrt{8\pi}\, x^{-4} \;, \, x \to \infty \;,
\end{cases}
\eeq
thus exhibiting a non-trivial algebraic tail $\propto x^{-4}$ for large $x$. Remarkably, based on numerical simulations, it was argued in Ref. \cite{UOSRW} that the scaling form in Eq. (\ref{eq:typ_scaling}) {\it and} the scaling function $P(x)$ are {\it universal}, i.e. they hold for any jump PDF $f(\eta)$ with a finite variance $\sigma^2$ [and with the substitution $\sigma_0 \to \sigma$ in Eq. (\ref{eq:typ_scaling})].

In Ref. \cite{UOSRW},  the universality of the gap density in Eq. (\ref{P_scaling}) was demonstrated for jump densities different from the symmetric exponential PDF, but only numerically. Hence, it would naturally be interesting to find other symmetric jump PDFs for which one can compute the limiting gap density analytically and demonstrate the universality. The most natural candidate for this is the so called symmetric Erlang density where      
\begin{equation}\label{eq:2}
\begin{split}
f_p\left(\eta\right) = \frac{|\eta|^p}{2p!b^{p+1}}e^{-\frac{|\eta|}{b}} \;, \qquad p\in \mathbb{N} \;.
\end{split}
\end{equation}
Without any loss of generality, and to simplify the computations, we set $b=1$ in the following. The case $p=0$ corresponds to the symmetric exponential density studied in \cite{UOSRW}. In this paper, we demonstrate that the same technique that was used for the $p=0$ in Ref. \cite{UOSRW} can be extended to other positive integer values of $p$. The Ref. \cite{UOSRW} was a short Letter and  the details of the computations could not
be presented. Here, we take this opportunity to present the technical details involved in this method. For the purpose of illustration, we first present the $p=0$ and the $p=1$ case so that the reader gets familiar with the technique (which is a bit involved) and then generalize to arbitrary positive integer $p$, rather than starting with the general $p$ and discuss special cases. Apart from this presentation of the technical details, the main result is to show that the same gap scaling function in Eq. (\ref{P_scaling}) also extends to Erlang jump densities for arbitrary integer shape parameter $p\in \mathbb{N}$.

{\bf Summary of main results}. It is useful to summarize our main results. For any jump PDF $f_p(\eta)$ (\ref{eq:2}) with non negative integer value $p$, we show explicitly that the PDF of the gaps $P_{k,n}(\Delta)$ reaches a limiting density $p_k(\Delta)$, as in Eq. (\ref{limiting}). For large $k$, we then show that $p_k(\Delta)$ exhibits two distinct regimes depending on the scale of $\Delta$: (i) $\Delta = O(1/\sqrt{k})$ corresponding to the {\it typical} fluctuations of the gaps and (ii) $\Delta = O(1)$ which corresponds to {\it large deviations} of the gaps -- and which were not studied in detail even for $p=0$ (\ref{def_Laplace}) in \cite{UOSRW}. These different behaviors can be summarized as follows
\bea\label{summary}
p_k(\Delta) \sim
\begin{cases}
&\dfrac{\sqrt{k}}{\sigma_p} P\left(\dfrac{\sqrt{k}}{\sigma_p} \Delta \right) \;, \; \Delta =  O(1/\sqrt{k}) \\
& \\
& \dfrac{1}{k^{3/2}} \varphi_p(\Delta) \;, \hspace*{0.65cm}\; \Delta =  O(1) \;,
\end{cases}
\eea
where $\sigma_p^2 = {(p+1)(p+2)}$ is the variance of the jump PDF $f_p(\eta)$ in (\ref{eq:2}) and $P(x)$ is the same density as for $p=0$ in Eq. (\ref{P_scaling}). This demonstrates that $P(x)$ actually holds for the symmetric Erlang jump densities for arbitrary parameter $p \in {\mathbb N}$. This is in agreement the conjecture about the universality of $P(x)$ in Ref. \cite{UOSRW} for symmetric jump densities with finite variance. 

Although less universal, the large deviation regime, for $\Delta = O(1)$ in Eq. (\ref{summary}), is also interesting. In fact, the $k$-dependence $\propto k^{-3/2}$ in the prefactor of the second line in Eq. (\ref{summary}) is universal and holds for any value of $p$. However, the function $\varphi_p(\Delta)$ is non universal. We can compute it explicitly for $p=0$ [see Eq. (\ref{phi_0})] and already for $p=1$ it has a quite complicated expression (and hence not reported here). Nevertheless, one can compute the asymptotic behaviors of $\varphi_p(\Delta)$ for arbitrary integer $p$. They are given by 
\bea\label{asympt_phip}
\varphi_p(\Delta) \sim
\begin{cases}
&\dfrac{3 \, \sigma_p^3}{\sqrt{8 \pi}}\,\Delta^{-4} \;, \; \Delta \to 0 \\
& \\
&C_p \, \Delta^{2p} e^{-2 \Delta} \;, \; \Delta \to \infty \;,
\end{cases}
\eea
where $C_p$ is a nontrivial constant that we have only obtained for $p=0$ (\ref{asympt_phi0}) and $p=1$ (\ref{C1}). Interestingly, by using the asymptotic behavior of the scaling function $P(x)$ for large $x$ (\ref{asympt_P}) and the one of $\varphi_p(\Delta)$ for small $\Delta$ (\ref{asympt_phip}), one can explicitly check that the typical regime, for $\Delta = O(1/\sqrt{k})$, and the large deviation regime, for $\Delta = O(1)$, of $p_k(\Delta)$ match smoothly. To see this, we first inject the large $x$ behavior of $P(x) \sim 3/\sqrt{8 \pi}x^{-4}$ (\ref{asympt_P}) in the first line of Eq. (\ref{summary}). This yields 
\beq\label{matching_1}
p_k(\Delta) \sim \frac{3}{\sqrt{8 \pi}} \left( \frac{\sigma_p}{\sqrt{k}}\right)^3 \Delta^{-4} = \frac{1}{k^{3/2}} \frac{3 \, \sigma_p^3}{\sqrt{8 \pi}} \, \Delta^{-4} \;, \, \Delta \gg \sigma_p/\sqrt{k}\;. 
\eeq
On the other hand, if one starts in the large deviation regime for $\Delta = O(1)$ and look at the limit $\Delta \to 0$, one can insert the small $\Delta$ behavior of $\varphi_p(\Delta) \sim (3 \sigma_p^3)/\sqrt{8 \pi} \Delta^{-4}$ [see Eq. (\ref{asympt_phip})] in the second line of Eq. (\ref{summary}) to obtain
\beq\label{matching_2}
p_k(\Delta) \sim \frac{1}{k^{3/2}} \frac{3 \, \sigma_p^3}{\sqrt{8 \pi}} \Delta^{-4}\;. 
\eeq  
By comparing Eq. (\ref{matching_1}) and (\ref{matching_2}), we clearly see that there is a smooth matching between the two regimes.

{\bf Outline.} The paper is organized as follows. In section 2, we describe the general framework, which is based on recursion relations and generating functions techniques, to study this problem. In section 3, we recall the exact solution obtained in Ref. \cite{UOSRW} for $p=0$ and in section 4 we extend this method to $p=1$. The exact solution for $p=1$ is quite cumbersome and instead we develop an asymptotic method to study the PDF of the gaps $P_{k,n}(\Delta)$ (\ref{def_pkn}) directly in the large $n$ and large $k$ limit. In section 5, we show how this asymptotic analysis can be carried out for any integer $p$, before we conclude in section 6. Some technical details have been relegated in appendices.

\section{General framework: recursion relations and generating functions}

In this section, we present a general framework, introduced in Ref. \cite{UOSRW}, to study the order statistics of RWs evolving according to (\ref{eq:1}). It is based on backward equations and is valid in principle for any symmetric jump PDF $f_p(\eta)$. We first focus on the PDF of $k$-th maximum $M_{k,n}$ and then on the gap $d_{k,n}$.

\subsection{Distribution of the $k$-th maximum $M_{k,n}$}

\begin{figure}
 \includegraphics[width=\linewidth]{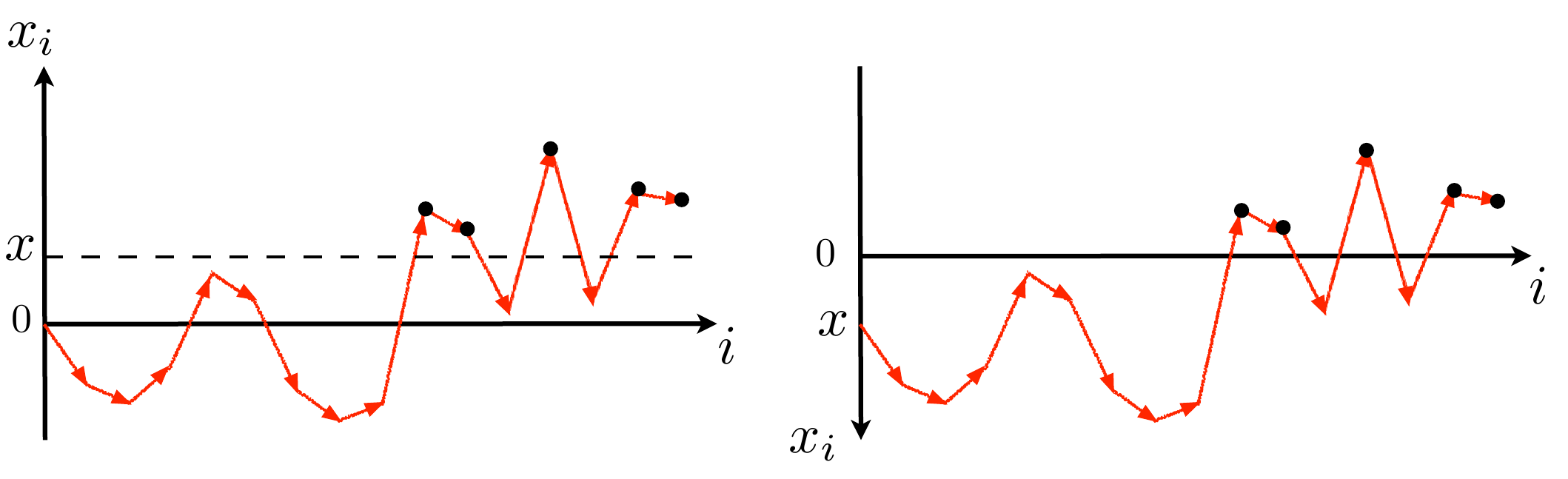}
 \caption{{\bf Left:} Realization of a RW with $n=15$ steps such that $5$ points are above $x \geq 0$ between step $1$ and step $n=15$. It contributes to the probability that $M_{k,15} \leq x$ for all $k \geq 6$. {\bf Right:} The same realization of the RW with $n=15$ steps as in the right panel but where the direction of the $y$-axis has been reversed. The lower half of the $y$-axis corresponds to positive values. Because the jump density is symmetric (\ref{eq:2}), this realization also contributes to the probability $q_{5,15}(x)$ that the RW, starting at $x$, has 5 points below the negative axis between step $1$ and step $n=15$.}\label{fig:illustration}
\end{figure}

The key quantity that leads to an explicit solution for the order statistics is the cumulative distribution function (CDF) of the $k$-th maximum, $F_{k,n}(x) = {\rm Pr}(M_{k,n}\leq x)$, for a walk starting at $x=0$. We want to derive a recursion relation for $F_{k,n}(x)$. A useful observation is that there is a one-to-one mapping between the event $M_{k,n} \leq x$ and the event that at most $k-1$ points of the RW are above $x$ between step $1$ and step $n$ (see Fig. \ref{fig:illustration}). Using the fact the jump density $f_p(\eta)$ in Eq. (\ref{eq:2}) is symmetric, i.e., $f_p(\eta) = f_p(-\eta)$, one can equivalently consider 
the quantity $q_{k,n}(x)$ which is the probability that the RW starting at $x_0=x$ has exactly $k$ points on the negative axis between step $1$ and step $n$ (see Fig. \ref{fig:illustration}). The CDF $F_{k,n}(x)$ can then be written in terms of $q_{k,n}(x)$ as
\begin{eqnarray}\label{CDF_max}
F_{k,n}(x) = 
\begin{cases}
&\sum_{m=0}^{k-1} q_{m,n}(x) \;, \; \hspace*{0.6cm}x \geq 0 \\
&\\
&\sum_{m=0}^{k-2} q_{n-m,n}(-x) \;, \; x \leq 0 \;,
\end{cases}
\end{eqnarray}  
where we used that $f_p(\eta)$ is symmetric and continuous. Note that even though the argument $x$ appearing in the function $F_{k,n}(x)$ can be negative, from Eq. (\ref{CDF_max}) it is clear that we need the function $q_{m,n}(x)$ only for positive argument $x\geq 0$. In all our subsequent calculations, we will assume that the argument $x\geq 0$ in $q_{k,n}(x)$.

The next step is to write a backward recursion equation for $q_{k,n}(x)$, by considering the jump $x \to x'$ at the first step and then using the Markov property of the RW in (\ref{eq:1}). One obtains, for~$n \geq 1$
\begin{eqnarray}\label{back_fp1_supp}
q_{k,n}(x) &=& \int_0^\infty q_{k,n-1}(x') f_p(x'-x) \, dx' + \int_{-\infty}^0 q_{n-k,n-1}(-x') f_p(x'-x) dx'  \;,
\end{eqnarray}
starting from $q_{0,0}(x) = 1$. The first term corresponds to a jump from $x$ to $x'>0$ while the second one corresponds to a jump from $x$ to $x'<0$. Note that, in the second term, the argument of $q_{n-k,n-1}(-x')$ is positive. To solve this equation (\ref{back_fp1_supp}) it is useful to introduce the auxiliary variable $r_{k,n}(x) \equiv q_{n-k,n}(x)$ which is  the probability that the RW defined above has $k$ points above $0$ between step $1$ and step $n$. One can then write two coupled equations for $q_{k,n}(x)$ and $r_{k,n}(x)$:
\begin{eqnarray}\label{back_fp1}
q_{k,n}(x) &=& \int_0^\infty q_{k,n-1}(x') f_p(x'-x) \, dx' + \int_0^\infty r_{k-1,n-1}(x') f_p(x'+x) dx'  \;,
\end{eqnarray}
which is directly obtained from (\ref{back_fp1_supp}) where we have simply made the change of variable $x' \to -x'$ and used that $f_p(\eta) = f_p(-\eta)$, and similarly
\begin{eqnarray}\label{back_fp2}
r_{k,n}(x) &=& \int_0^\infty r_{k-1,n-1}(x') f_p(x'-x) \, dx' + \int_0^\infty q_{k,n-1}(x') f_p(x'+x) \, dx' \;,
\end{eqnarray}
together with the initial conditions $q_{0,0}(x) = r_{0,0}(x) = 1$ and the convention that $q_{k,n}(x) = r_{k,n}(x)=0$ if $k > n$. To study these equations (\ref{back_fp1}) and (\ref{back_fp2}) it is natural to introduce the double generating functions (GF), for $s<1$ and $z<1$
\begin{eqnarray}\label{def_GFs}
\tilde q(z,s; x) = \sum_{n=0}^\infty \sum_{k=0}^n s^n z^k q_{k,n}(x) \;, \; \tilde r(z,s; x) = \sum_{n=0}^\infty \sum_{k=0}^n s^n z^k r_{k,n}(x) \;.
\end{eqnarray}
Using the fact that $r_{k,n}(x) = q_{n-k,n}(x)$, it is easy to see that $\tilde q(z,s;x)$ and $\tilde r(z,s;x)$ are related via the relation
\begin{eqnarray}\label{rel_q_r}
\tilde r(z,s;x) = \tilde q(1/z,z\,s;x) \;.
\end{eqnarray}
Finally, from Eqs. (\ref{back_fp1}) and (\ref{back_fp2}) one obtains that these functions $\tilde q(z,s;x)$ and $\tilde r(z,s;x)$ satisfy the coupled set of equations
\begin{eqnarray}
&&\hspace*{-0.5cm}\tilde q(z,s;x) = 1 + s \int_0^\infty f_p(x'-x) \tilde q(z,s;x') \, dx'  + zs \int_0^\infty f_p(x+x') \tilde r (z,s;x') \, dx' \label{fr_gf1} \\
 &&\hspace*{-0.5cm}\tilde r(z,s;x) = 1 + zs \int_0^\infty  f_p(x-x') \tilde r(z,s;x') \, dx' + s \int_0^\infty f_p(x+x') \tilde q(z,s;x') \, dx' \;. \label{fr_gf2}
\end{eqnarray}

\subsection{Distribution of the $k$-th gap}

One can then follow a similar strategy \cite{UOSRW} to compute the PDF of the $k$-th gap $d_{k,n} = M_{k,n} - M_{k+1,n}$, denoted as $P_{k,n}(\Delta)$ [see Eq. (\ref{def_pkn})]. For this we introduce the joint CDF $S_{k,n}(x,y) = {\rm Pr}[M_{k,n} > y, M_{k+1,n}<x]$ with $y>x$. Knowing this CDF $S_{k,n}(x,y)$, one can then compute the PDF $P_{k,n}(\Delta)$ from the relation
\begin{equation}\label{start_pdf_delta}
{P}_{k,n}(\Delta) =  - \int_{\mathbb{R}^2} \frac{\partial^2 {S}_{k,n}(x,y)}{\partial x \partial y} \theta(y-x) \theta(x + \Delta - y) dx dy \;.
\end{equation}
To compute $S_{k,n}(x,y)$ we introduce the probability ${Q}_{k,n}(x,\Delta)$ which is the probability that a RW of $n$ steps, defined by Eq. (\ref{eq:1}) and starting from $x_0 = x$, has $k$ points in the interval $(-\infty, -\Delta]$ (with $k \geq 1$) and $n-k$ points in the interval $[0, +\infty)$, hence with no point in the interval $[-\Delta, 0]$, from step $1$ to step $n$. The joint cumulative PDF $S_{k,n}(x,y)$ can be related to $Q_{k,n}$ as
\begin{eqnarray}\label{start_2}
{S}_{k,n}(x,y) = 
\begin{cases}
&{Q}_{k,n}(x, y-x) \;, \; x > 0 \\ 
& 0 \;, \; x < 0 \;\; {\rm and} \;\; y > 0 \\ 
& {Q}_{n-k+1,n}(-y,y-x) \;,\; x < 0 \;\; {\rm and} \;\; y < 0 \;.
\end{cases}
\end{eqnarray}
We note once again that we only need the function $Q_{k,n}(x,\Delta)$ for both $x\geq 0$ and $\Delta \geq 0$. In all the subsequent analysis, we will assume that both arguments of $Q_{k,n}(x,\Delta)$ are positive. Similarly to Eq. (\ref{back_fp1_supp}) one can derive a backward equation for ${Q}_{k,n}(x, \Delta)$. It reads \cite{UOSRW}, for $n \geq 1$:
\begin{eqnarray}\label{back_fptilde1_supp}
\hspace*{-2cm}{Q}_{k,n}(x, \Delta) &=& \int_0^\infty {Q}_{k,n-1}(x', \Delta) f_p(x-x') \, dx'  + \int_{-\infty}^0 {Q}_{n-k,n-1}(-x',\Delta) f_p(x-x'+\Delta) dx'  \;, \nonumber
\end{eqnarray}   
starting from ${Q}_{0,0}(x, \Delta)=1$. As done above for the $k$-th maximum, it is convenient to introduce an auxiliary quantity, ${R}_{k,n}(x, \Delta) = Q_{n-k,n}(x, \Delta)$, which is the probability that the RW has $k$ points above $0$ between step $1$ and step $n$, and $n-k$ points below $-\Delta$. These two quantities $Q_{k,n}(x, \Delta)$ and $R_{k,n}(x,\Delta)$ satisfy the following set of two coupled backward equations:
\begin{eqnarray}\label{back_fptilde1}
\hspace*{-2cm}{Q}_{k,n}(x, \Delta) = \int_0^\infty {Q}_{k,n-1}(x', \Delta) f_p(x-x') \, dx'  + \int_0^\infty {R}_{k-1,n-1}(x',\Delta) f_p(x'+x+\Delta) dx'  \;, 
\end{eqnarray}   
which is obtained from (\ref{back_fptilde1_supp}) with the change of variable $x' \to -x'$ and similarly
\begin{eqnarray}\label{back_fptilde2}
\hspace*{-2cm}{R}_{k,n}(x, \Delta) &=& \int_0^\infty {R}_{k-1,n-1}(x', \Delta) f_p(x'-x) \, dx'  + \int_0^\infty {Q}_{k,n-1}(x',\Delta) f_p(x'+x + \Delta) \, dx', 
\end{eqnarray}
together with the initial conditions ${Q}_{0,0}(x, \Delta) = {R}_{0,0}(x, \Delta) = 1$ and, again, with the convention that ${Q}_{k,n}(x, \Delta) = {R}_{k,n}(x, \Delta)=0$ if $k > n$. To study these recurrence equations for $Q_{k,n}(x,\Delta)$ and $R_{k,n}(x,\Delta)$ (\ref{back_fptilde1}) and (\ref{back_fptilde2}), it is convenient to introduce their double generating functions
\begin{eqnarray}\label{expl_qcal1}
&&\tilde {Q}(z,s;x, \Delta) = \sum_{n=0}^\infty \sum_{k=0}^n s^n z^k  {Q}_{k,n}(x,\Delta) \;, \label{expl_qcal1} \\
&&\; \tilde {R}(z,s;x, \Delta) = \sum_{n=0}^\infty \sum_{k=0}^n s^n z^k {R}_{k,n}(x,\Delta) \;. \label{expl_pcal1}
\end{eqnarray}
As above (\ref{rel_q_r}), the relation $R_{k,n}(x,\Delta) = Q_{n-k,n}(x,\Delta)$ implies the following relation between their GF
\begin{eqnarray}
\tilde {R}(z,s;x, \Delta) = \tilde Q(1/z,z\,s;x,\Delta) \;.
\end{eqnarray}
Finally, from the coupled recurrence relations in Eqs. (\ref{back_fptilde1}) and (\ref{back_fptilde2}), one obtains that the GFs satisfy the coupled set of integral equations
\begin{equation}\label{eq:WH}
\begin{split}
\tilde{Q}(z,s;x,\Delta) &= 1+s\int_{0}^{\infty}\tilde{Q}(z,s;x',\Delta)f_p(x-x')dx' + zs\int_{0}^{\infty}\tilde{R}(z,s;x',\Delta)f_p(x'+x+\Delta)dx'\\
\tilde{R}(z,s;x,\Delta) &= 1+zs\int_{0}^{\infty}\tilde{R}(z,s;x',\Delta)f_p(x-x')dx'+ s\int_{0}^{\infty}\tilde{Q}(z,s;x',\Delta)f_p(x'+x+\Delta)dx' \;,
\end{split}
\end{equation}
where we recall that $x \geq 0$ as well as $\Delta \geq 0$. Once these equations (\ref{eq:WH}) for $\tilde{Q}(z,s;x,\Delta)$ and $\tilde{R}(z,s;x,\Delta)$ are solved, the PDF of the gap $P_{k,n}(\Delta)$ can be computed from Eqs. (\ref{start_pdf_delta}) and (\ref{start_2}). This yields
\begin{equation}\label{eq:kgap}
 \begin{split}
P_{k,n}(\Delta) = &-\int_{0}^{\infty}dx\int_{x}^{\infty}dy\frac{\partial^2}{\partial x\partial y}Q_{k,n}(x,y-x)\delta(y-x-\Delta)\\ &- \int_{-\infty}^{0}dx\int_{x}^{0}dy\frac{\partial^2}{\partial x\partial y}R_{k-1,n}(-y,y-x)\delta(y-x-\Delta) \;,
 \end{split}
 \end{equation} 
where we have also used $R_{k,n}\left(x,\Delta\right) = Q_{n-k,n}\left(x,\Delta\right)$. If one introduces the double GF of $P_{k,n}(\Delta)$
\begin{eqnarray}\label{def_ptilde}
\tilde P(z,s;\Delta) = \sum_{n=0}^\infty \sum_{k=0}^\infty s^n z^k P_{k,n}(\Delta) \;,
\end{eqnarray}
one immediately obtains from Eq. (\ref{eq:kgap}) that it satisfies
\begin{equation}\label{eq:6}
\begin{split}
\tilde{P}(z,s;\Delta) =& -\int_{0}^{\infty}dx\int_{x}^{\infty}dy\frac{\partial^2}{\partial x\partial y}\tilde Q(z,s;x,y-x)\delta(y-x-\Delta)\\ &- z\,\int_{-\infty}^{0}dx\int_{x}^{0}dy\frac{\partial^2}{\partial x\partial y}\tilde R(z,s;-y,y-x)\delta(y-x-\Delta) \;.
\end{split} 
\end{equation}
The goal is thus quite clear: we have to solve this set of equations for $\tilde Q$ and $\tilde R$ in (\ref{eq:WH}) and then insert these solutions in Eq. (\ref{eq:6}) to compute the GF of the PDF of the gaps from Eq. (\ref{eq:6}). However, these integral equations for $\tilde Q$ and $\tilde R$ in Eq. (\ref{eq:WH}) are of the Wiener-Hopf type and they are very hard to solve for generic jump density. In Ref. \cite{UOSRW} it was shown that they can be solved exactly for the $p=0$ case, i.e., for $f_0(\eta) = \frac{1}{2} e^{-|\eta|}$. Indeed in this case, these integral equations can be transformed
into differential equations, which can eventually be solved exactly. The goal of this paper is to show that we can actually extend the method of Ref. \cite{UOSRW} to a wider class of jump PDFs $f_p(\eta) \propto |\eta|^p\, e^{-|\eta|}$ (\ref{eq:2}) with $p$ being a non-negative integer. Of course, the most interesting features of the PDF of the gap $P_{k,n}(\Delta)$ emerge in the limit where both $n$ and $k$ are large. This means that we will be mainly interested in the behavior when $s \to 1$ (corresponding to $n \to \infty$) and $z \to 1$ (corresponding to $k \to \infty$) of the GF $\tilde Q(z,s;x,\Delta)$ and $\tilde R(z,s;x,\Delta)$. 

Below, in section 3, we first recall the results obtained for $p=0$, where both $\tilde Q(z,s;x,\Delta)$ and $\tilde R(z,s;x,\Delta)$ can be computed explicitly for all $0<s<1$ and $0<z<1$, and from which the asymptotic behaviors for $s \to 1$ and $z \to 1$ can be easily obtained. In section 4 we will see, in the case $p=1$, how to obtain directly the asymptotic behaviors of  
$\tilde Q(z,s;x,\Delta)$ and $\tilde R(z,s;x,\Delta)$ in that limit from Eqs. (\ref{eq:WH}). We will then extend this asymptotic analysis to arbitrary integer values of $p$ in section 5. 


\section{The case of the symmetric exponential jump density ($p=0$)}

We first consider the case where the PDF of increments is given by 
\begin{eqnarray}\label{def_exr_inc}
f_0(x) = \frac{1}{2} \exp{(-|x|)} \;.
\end{eqnarray} 
For completeness, we recall the main steps that led to the
exact solution for the PDF of the gap $P_{k,n}(\Delta)$ \cite{UOSRW}. A key property
of this density (\ref{def_exr_inc}) is that it satisfies the identity
\begin{eqnarray}\label{sec_deriv_exp}
f_0'(x) = - \frac{1}{2} \sgn{(x)} \exp{(-|x|)} \;, \; f_0''(x) = -\delta(x) + f_0(x) \;.
\end{eqnarray}
Note that the second relation in Eq. (\ref{sec_deriv_exp}) can also be written as
\begin{eqnarray}\label{sec_deriv_exp_2}
(1-D^2)f_0(x) = \delta(x) \;,
\end{eqnarray}
where $D = \, \frac{d}{dx}$, a relation that can be generalized to other jump PDFs of the type given in Eq. (\ref{eq:2}). We start with the statistics of the $k$-th maximum $M_{k,n}$, which allows us to compute the first moment of the gaps $\langle d_{k,n} \rangle= \langle M_{k,n}\rangle - \langle M_{k+1,n}\rangle$. The study of the full PDF of the gaps will be carried out in a second step.

\subsection{Statistics of the $k$-th maximum}

Using the identity in Eq. (\ref{sec_deriv_exp}), one can take the second derivative with respect to $x$ of the equations (\ref{back_fp1}), (\ref{back_fp2}) to obtain
\begin{eqnarray}\label{pde}
\begin{split}
&&\frac{\partial^2 q_{k,n}(x)}{\partial x^2} = -  q_{k,n-1}(x) +  q_{k,n}(x) \;, \\
 &&\frac{\partial^2 r_{k,n}(x)}{\partial x^2} =  -  r_{k-1,n-1}(x) + r_{k,n}(x) \;.
 \end{split}
\end{eqnarray}
One thus finds that, for $p=0$, the coupled integral equations yield two independent differential recurrence equations (\ref{pde}). They can be solved by introducing the double GFs, for $s<1, z<1$:
\begin{eqnarray}
\tilde q(z,s; x) = \sum_{n=0}^\infty \sum_{k=0}^n s^n z^k q_{k,n}(x) \;, \; \tilde r(z,s; x) = \sum_{n=0}^\infty \sum_{k=0}^n s^n z^k r_{k,n}(x) \;,
\end{eqnarray}
which satisfy the following equations:
\begin{eqnarray}
&&\frac{\partial^2 \tilde q(z,s;x)}{\partial x^2} = (1-s) \tilde q(z,s;x) - 1 \;, \label{eq_q1}\\
&&\frac{\partial^2 \tilde r(z,s;x)}{\partial x^2} = (1-zs) \tilde r(z,s;x) - 1 \;. \label{eq_r1}
\end{eqnarray}
We recall that these equations are valid for $x \geq 0$. The only boundary conditions that fix the solution uniquely are that as $x \to +\infty$, both $\tilde q(z,s;x)$ and $\tilde r(z,s;x)$ should not diverge. The boundary conditions at $x=0$ are not specified, but will be determined by the equations themselves. Let us first consider the equation for $\tilde q(z,s;x)$ in (\ref{eq_q1}). The general solution is of the form
\begin{eqnarray}
\tilde q(z,s;x) =  \frac{1}{1-s}  + a(z,s) \exp{\left(-t_1(s) \, {x}\right)} + \tilde a(z,s) \exp{\left(-\tilde t_1(s)\,{x}\right)}
\end{eqnarray}
where $t_1(s)$ and $\tilde t_1(s)$ are the two solutions of 
\begin{eqnarray}\label{eq_lambda_exp}
t^2 = 1- s  \Longrightarrow t_1(s) = -\tilde t_1(s) = \sqrt{1-s} \;.
\end{eqnarray}
Since the solution for $\tilde q(z,s;x)$ can not diverge as $x \to \infty$, we discard the negative root, i.e. set $\tilde a = 0$,  and keep only the positive root. This gives 
\begin{eqnarray}\label{expl_q1}
\hspace*{-2cm}\tilde q(z,s;x) = a(z,s) \exp{\left(-t_1(s) {x}\right)} + \frac{1}{1-s}  =  a(z,s) \exp{\left(-\sqrt{1-s} \, {x}\right)} + \frac{1}{1-s}\;.
\end{eqnarray}
Similarly the solution of Eq. (\ref{eq_r1}) for $\tilde r(z,s;x)$ reads
\begin{eqnarray}\label{expl_p1}
\hspace*{-2cm} \tilde r(z,s;x) = b(z,s) \exp{\left(-t_1(zs) {x}\right)} + \frac{1}{1-z \, s}= b(z,s) \exp{\left(-\sqrt{1-zs} \, {x}\right)} + \frac{1}{1-z \, s} \;,
\end{eqnarray}
where, at this stage, the amplitudes $a(z,s)$ and $b(z,s)$ remain undetermined. To determine them, we substitute back these forms (\ref{expl_q1}) and (\ref{expl_p1}) into the integral equations satisfied by $\tilde q(z,s;x)$ and $\tilde r(z,s;x)$ in Eqs. (\ref{fr_gf1}) and (\ref{fr_gf2}). One thus obtains a linear system of coupled equations for $a(z,s)$ and $b(z,s)$ which reads
\begin{eqnarray}
&& 0 = - \frac{1}{1-s} + \frac{a(z,s)}{\sqrt{1-s}-1} + \frac{z}{1-z\,s} + \frac{z \, b(z,s)}{1+\sqrt{1-zs}} \label{lin_ab1}\\
&&0 = - \frac{z}{1-zs} + \frac{z \, b(z,s)}{\sqrt{1-z\,s}-1} + \frac{1}{1-s}  + \frac{a(z,s)}{1+\sqrt{1-s}} \label{lin_ab2}\;.
\end{eqnarray}
 This system can be straightforwardly solved, yielding
 \begin{eqnarray}\label{explicit_ab}
 &&a(z,s) = \frac{1}{\sqrt{1-s}\sqrt{1-zs}} - \frac{1}{1-s} \\
 &&b(z,s) = \frac{1}{\sqrt{1-s}\sqrt{1-zs}} - \frac{1}{1-zs} \;.
 \end{eqnarray}
Note that one explicitly checks that $b(z,s) = a(1/z,z s)$, as it should [see Eq. (\ref{rel_q_r})]. One obtains finally an explicit formula for $\tilde q(z,s;x)$ 
\begin{eqnarray}\label{final_explicit_Q}
\tilde q(z,s;x) =  \frac{1}{1-s} +\left(\frac{1}{\sqrt{(1-s)(1-zs)}} - \frac{1}{1-s} \right)\exp{\left(-\sqrt{2(1-s)} \, \frac{x}{\sigma_0}\right)}  \;, \nonumber
\end{eqnarray}
in terms of the standard deviation $\sigma_0$ which reads
\begin{eqnarray}
\sigma_0 = \sqrt{2} \;. \label{sigma_exp}
\end{eqnarray}
Using the fact that the CDF of the $k$-th maximum $F_{k,n}(x) = {\rm Pr}[M_{k,n} \leq x]$ can be expressed as a linear combination of $q_{m,n}(x)$ [see Eq. (\ref{CDF_max})], we now compute the double GF of the moments $\langle M_{n,k}^m\rangle$ in the exponential case from the explicit expression above (\ref{final_explicit_Q}). Let us consider here the first moment in detail. From Eq. (\ref{CDF_max}) one has
 \begin{eqnarray}\label{mkn_inter}
 \langle M_{k,n} \rangle = \sum_{m=0}^{k-1} \int_0^\infty x \frac{\partial}{\partial x} q_{m,n}(x) \, dx + \sum_{m=0}^{k-2} \int_0^\infty x \frac{\partial}{\partial x} q_{n-m,n}(x) \, dx \;.
 \end{eqnarray}
 One can then compute the following GF:
 \begin{eqnarray}
 &&\sum_{n=0}^\infty \sum_{m=0}^n z^m s^n \int_0^\infty x \frac{\partial}{\partial x} q_{m,n}(x) \, dx  = \frac{\sigma_0}{\sqrt{2}} \left(\frac{1}{(1-s)^{3/2}} - \frac{1}{(1-s)}\frac{1}{\sqrt{1-zs}} \right) \\
 && = \frac{\sigma_0}{\sqrt{2}} \left(\frac{2}{\sqrt{\pi}} \sum_{n=0}^\infty \frac{\Gamma(n+3/2)}{n+1} s^n - \frac{1}{\sqrt{\pi}} \sum_{n=0}^\infty s^n \sum_{m=0}^n z^m \frac{\Gamma(m+1/2)}{\Gamma(m+1)}   \right) \;,
 \end{eqnarray}
from which one obtains
\begin{eqnarray}\label{final_intqprime}
\int_0^\infty x \frac{\partial}{\partial x} q_{m,n}(x) \, dx = 
\begin{cases}
&\dfrac{\sigma_0}{\sqrt{2}} \left(\dfrac{2}{\sqrt{\pi}} \dfrac{\Gamma(n+3/2)}{\Gamma(n+1)} - 1 \right) \;, \; m = 0 \\
&\\
&-\dfrac{\sigma_0}{\sqrt{2\pi}} \dfrac{\Gamma(m+1/2)}{\Gamma(m+1)} \;, \hspace*{1.2cm}\;m >0 \;.
\end{cases}
\end{eqnarray}
Finally plugging these expressions (\ref{final_intqprime}) in the formula for $\langle M_{k,n}\rangle$ in (\ref{mkn_inter}), one finds
\begin{eqnarray}
\langle M_{k,n}\rangle = \frac{\sigma_0}{\sqrt{2 \pi}} \sum_{m=k}^{n-k+1} \frac{\Gamma(m+1/2)}{\Gamma(m+1)} \;.
\end{eqnarray}
Note that we have used the formula (which can be computed using {\it Mathematica} and can be easily shown by the method of induction)
\begin{eqnarray}
\sum_{m=0}^n \frac{\Gamma(m+1/2)}{\Gamma(m+1)} = 2 \frac{\Gamma(n+3/2)}{\Gamma(n+1)} \;.
\end{eqnarray}
It is then straightforward to extract the first moment of the gap $\langle d_{k,n}\rangle$ as
\begin{eqnarray}\label{exact_mean_gar_exp}
\langle d_{k,n} \rangle = \sigma_0 \left(\frac{\Gamma\left(k + \frac{1}{2} \right)}{ \sqrt{2\pi} k !} + \frac{\Gamma\left(n-k + \frac{3}{2} \right)}{ \sqrt{2\pi} (n-k) !} \right)\;,
\end{eqnarray}
which depends in a non-trivial way both on $k$ and $n$. Note also that one can check explicitly that 
$\langle d_{k,n} \rangle = \langle d_{n-k+1,n} \rangle$, which reflects the symmetry of the RW (\ref{eq:1}) under the transformation $x \to - x$ for symmetric jump PDFs considered here. From Eq. (\ref{exact_mean_gar_exp}) one easily obtains that, in the limit of large $n$
\beq\label{gap_largen}
\lim_{n \to \infty} \frac{\langle d_{k,n}\rangle}{\sigma_0} = \frac{\Gamma(k+1/2)}{\sqrt{2 \pi} \, k!} \;.
\eeq
In addition, in the limit of large $k$, one finds \cite{UOSRW}
\beq\label{gap_largenk}
\lim_{n \to \infty} \frac{\langle d_{k,n}\rangle}{\sigma_0} = \frac{1}{\sqrt{2 \pi k}} + O(k^{-1}) \;,
\eeq
a formula that can also be obtained from the so-called Pollaczek-Wendel identity known in fluctuation theory \cite{FP1952,wendel}, as shown in Ref. \cite{UOSRW}.

\subsection{Statistics of the $k$-th gap}

We now come to the PDF of the gap $d_{k,n}$. The GF of the gap probability $P_{k,n}(\Delta)$ is obtained from Eq. (\ref{eq:6}) in terms of $\tilde Q(z,s;x, \Delta)$ and $\tilde R(z,s;x,\Delta)$ which are solutions of Eq.~(\ref{eq:WH}).

In the case of a asymmetric exponential density of the increments, $f_0(x) = \frac{1}{2} \exp{(-|x|)}$, one can use the properties mentioned above (\ref{sec_deriv_exp}) to obtain that ${Q}_{k,n}(x,\Delta)$ and ${R}_{k,n}(x,\Delta)$ satisfy exactly the same differential recurrence equations as ${q}_{k,n}(x)$ and ${r}_{k,n}(x)$ in Eqs. (\ref{pde}). Hence their double GF are also of the form
\begin{eqnarray}\label{expl_qcal1}
\hspace*{-1.cm}\tilde {Q}(z,s;x, \Delta) = \sum_{n=0}^\infty \sum_{k=0}^n s^n z^k  {Q}_{k,n}(x,\Delta) &=& {A}_1(z,s; \Delta) \exp{\left(-t_1(s)\,{x}\right)} + \frac{1}{1-s} \\
&=& {A}_1(z,s; \Delta) \exp{\left(-\sqrt{1-s} \, {x}\right)} + \frac{1}{1-s} \;, 
\end{eqnarray}
where $t_1(s) = \sqrt{1-s}$ is the positive root of Eq. (\ref{eq_lambda_exp}) and similarly
\begin{eqnarray}\label{expl_pcal1}
\hspace*{-1.cm} \tilde {R}(z,s;x, \Delta) = \sum_{n=0}^\infty \sum_{k=0}^n s^n z^k {R}_{k,n}(x,\Delta) &=& {B}_1(z,s; \Delta) \exp{\left(-t_1(zs) {x}\right)} + \frac{1}{1-zs} \\
&=& {B}_1(z,s; \Delta) \exp{\left(-\sqrt{1-zs} \,{x}\right)} + \frac{1}{1-zs}\;.
\end{eqnarray}
However the amplitudes $A_1(z,s; \Delta)$ and ${B}_1(z,s; \Delta)$ satisfy a set of two linear equations which are different from Eqs. (\ref{lin_ab1}, \ref{lin_ab2}). They are obtained by injecting the above expressions (\ref{expl_qcal1}, \ref{expl_pcal1}) in the integral equations satisfied by $\tilde {Q}(z,s;x, \Delta)$ and $\tilde {R}(z,s;x, \Delta)$ in Eq. (\ref{eq:WH}). This yields a linear system of equation for ${A}_1(z,s; \Delta)$ and ${B}_1(z,s; \Delta)$:
\begin{eqnarray}
&& 0 = - \frac{1}{1-s} + \frac{{A}_1}{\sqrt{1-s}-1} + \frac{z}{1-zs} \exp{\left(-{\Delta}\right)} + \frac{z \, {B}_1}{1+\sqrt{1-zs}}\exp{\left(-{\Delta}\right)}  \label{lin_ab1}\\
&&0 = - \frac{z}{1-zs} + \frac{z {B}_1}{\sqrt{1-zs}-1} + \frac{1}{1-s}\exp{\left(-{\Delta}\right)}   + \frac{{A}_1}{1+\sqrt{1-s}}\exp{\left(-{\Delta}\right)} \label{lin_ab2}\;.
\end{eqnarray}
where ${A_1} \equiv {A_1}(z,s;\Delta)$ and similarly ${B_1} \equiv {B_1}(z,s;\Delta)$. Solving this linear system (\ref{lin_ab1}) and (\ref{lin_ab2}) for $A_1$ and $B_1$ yields 
\begin{eqnarray}
&&{A}_1(z,s; \Delta) = \frac{\frac{s}{1-s}(\sqrt{1-zs}+1) - 2 \frac{zs}{\sqrt{1-zs}}e^{-\Delta} + \frac{s}{1-s}(\sqrt{1-zs}-1) e^{-2\,\Delta}   }{e^{-2 \Delta}(\sqrt{1-s}-1)(\sqrt{1-zs}-1) - (\sqrt{1-s}+1)(\sqrt{1-zs}+1)} \label{exact_A1}\\
&& \nonumber \\
&&{B}_1(z,s; \Delta) = {A}_1(1/z, z\, s; \Delta) \;. \label{exact_B1}
\end{eqnarray}
From Eq. (\ref{eq:6}) one eventually obtains the GF of the PDF of the gap (see \ref{sec:app:64})
\begin{eqnarray}\label{GF_pdf_delta}
\hspace*{-2.5cm}\tilde {P}(z,s;\Delta) &=& \partial_{\Delta} {A}_1(z,s; \Delta) + \frac{1}{t_1(s)} \partial_\Delta^2 {A}_1(z,s; \Delta) \nonumber  \\
&& + z \,e^{t_1(zs)\Delta}\left(\partial_{\Delta} {B}_1(z,s; \Delta) + \frac{1}{t_1(zs)} \partial_\Delta^2 {B}_1(z,s; {\Delta})\right) \nonumber \\
\hspace*{-2.5cm}&=&\partial_{\Delta} {A}_1(z,s; \Delta) + \frac{1}{\sqrt{1-s}} \partial_\Delta^2 {A}_1(z,s; \Delta) \nonumber \\
&&+ z\,e^{\sqrt{1-zs} \, \Delta} \left(\partial_{\Delta} {B}_1(z,s; \Delta) + \frac{1}{\sqrt{1-zs}} \partial_\Delta^2 {B}_1(z,s; {\Delta})\right)  \;, 
\end{eqnarray}
where, in the third and fourth lines, we have used that $t_1(s) = \sqrt{1-s}$. This is an exact formula, from which in principle $P_{k,n}(\Delta)$ can be extracted, for any $k$ and $n$, by expanding $\tilde {P}(z,s;\Delta)$ in powers of $z$ and $s$ -- although obtaining an explicit formula for any $k$ and $n$ is still hard. However, we are mainly interested in the behavior of $P_{k,n}(\Delta)$ in the limit of a large number of steps, $n \gg 1$.

\subsection{Asymptotic analysis}

To study this large $n$ limit, one needs to analyse the behavior of the generating function $\tilde {P}(z,s;\Delta)$ in the limit $s \to 1$. The first important property to notice is that the (positive) root $t_1(s)$ vanishes as $s \to 1$ with a square-root singularity, $t_1(s) = \sqrt{1-s}$ [see Eq. (\ref{eq_lambda_exp})]. From the exact expressions in Eqs. (\ref{exact_A1}) and (\ref{exact_B1}) one extracts the following asymptotic behaviors as~$s \to 1$
\begin{eqnarray}
&&A_1(z,s; \Delta) \sim -\frac{1}{1-s} + \frac{A_{1,1}(z; \Delta)}{\sqrt{1-s}} + O((1-s)^0) \label{asympt_s1_A} \\
&&B_1(z,s;\Delta) \sim  \frac{B_{1,1}(z; \Delta)}{\sqrt{1-s}} + O((1-s)^0) \label{asympt_s1_B}
\end{eqnarray}
where 
\begin{equation}\label{A12}
\hspace*{-0.cm}A_{1,1}(z;\Delta) = \frac{\cosh(\Delta) + \sqrt{1-z}\sinh{(\Delta)}}{\sinh(\Delta) + \sqrt{1-z}\cosh{(\Delta)}} \;, \; B_{1,1}(z;\Delta) = \frac{1}{\sinh(\Delta) + \sqrt{1-z}\cosh{(\Delta)}} \;.
\end{equation}
If one inserts these expansions (\ref{asympt_s1_A}) and (\ref{asympt_s1_B}) in the general expression for the GF of the PDF of the gaps in Eq. (\ref{GF_pdf_delta}) one finds that the leading term when $s \to 1$ is the second term in Eq.~(\ref{GF_pdf_delta}),~i.e. 
\begin{eqnarray}\label{asympt_tilde_P}
\tilde P(z,s;\Delta) \sim \frac{1}{1-s} \partial^2_\Delta A_{1,1}(z;\Delta) + {\cal O}\left(\frac{1}{\sqrt{1-s}}\right) \;.
\end{eqnarray}
This behavior $\propto 1/(1-s)$ indicates that $P_{k,n}(\Delta)$ converges to a stationary density as $n \to \infty$, 
\begin{eqnarray}\label{stationary}
\lim_{n \to \infty} P_{k,n}(\Delta) = p_{k}(\Delta) \;, \;
\end{eqnarray} 
and the GF of $p_k(\Delta)$ reads
\begin{eqnarray}\label{expl_stationary}
 \tilde p(z;\Delta) = \sum_{k=1}^\infty z^k p_{k}(\Delta) = \partial^2_\Delta A_{1,1}(z;\Delta) = {8z}\, e^{-2 {\Delta}} \frac{u(z) - v(z) e^{-2 {\Delta}}}{[u(z) + v(z) e^{-2 {\Delta}}]^3}  \;,
\end{eqnarray} 
with $u(z) = \sqrt{1-z} + 1$ and $v(z) = \sqrt{1-z} - 1$.


\subsubsection{Typical fluctuations of the gap.}

Extracting $p_{k}(\Delta)$ from this GF (\ref{expl_stationary}) remains a difficult task. However, guided by our previous finding that the first moment $\langle d_{k,n} \rangle \sim k^{-1/2}$ for large $k$~(\ref{gap_largenk}), it is natural to study $\tilde p(z; \Delta)$ in the scaling regime where $(1-z) \to 0$ (corresponding to large $k$) and $\Delta \to 0$ keeping the ratio 
\bea\label{scaling_var}
\lambda = \frac{\sigma_0^2(1-z)}{\Delta^2}
\eea
fixed, where we recall that $\sigma_0 = \sqrt{2}$. In view of future computations, it is useful to analyze the behavior of $A_{1,1}$ and $B_{1,1}$ in Eq. (\ref{A12}) in that limit. One finds
\beq\label{scaling_form_p0}
A_{1,1} \sim \frac{1}{\sqrt{1-z}} a_1(\lambda) \;, \; B_{1,1} \sim \frac{1}{\sqrt{1-z}} b_1(\lambda)
\eeq
with 
\beq\label{explicit_ab_p0}
a_1(\lambda) = b_1(\lambda) = \frac{1}{1+\sqrt{2/\lambda}} \;.
\eeq
Using these results (\ref{scaling_form_p0}) and (\ref{explicit_ab_p0}) one obtains from Eq. (\ref{expl_stationary}) that 
$\tilde p(z; \Delta)$ takes the scaling form in this scaling limit
\bea\label{scaling_form_p0bis}
\tilde p(z; \Delta) \sim \frac{\sigma_0^2}{\Delta^3} \frac{1}{(1+\sqrt{\lambda/2})^3} \;.
\eea 
This scaling form for the generating function $\tilde p(z; \Delta)$ implies that the PDF $p_k(\Delta)$ is of the form, in the limit where $k \gg 1$, $\Delta \to 0$, keeping $\sqrt{k} \, \Delta$ fixed,
\begin{eqnarray}\label{scaling_form}
p_{k}(\Delta) \sim \frac{1}{\sigma_0}\sqrt{k} \, P\left(\sqrt{k} \frac{\Delta}{\sigma_0}\right) \;, 
\end{eqnarray}
where the function $P$ is independent of $k$. One checks that this scaling form (\ref{scaling_form}) is indeed compatible with the expression for the GF above (\ref{scaling_form_p0}) which in addition provides the following relation
\begin{eqnarray}\label{laplace_tf}
\int_0^\infty e^{- x \lambda} \sqrt{x} P(\sqrt{x}) \, dx = \frac{1}{(1 + \sqrt{\lambda/2})^3}  \;.
\end{eqnarray}
This Laplace transform above (\ref{laplace_tf}) can be inverted by using the following integral representation
\begin{eqnarray}\label{trick}
\frac{1}{(1 + \sqrt{\lambda/2})^3} = \frac{1}{2} \int_0^\infty  y^2 e^{-y - y \sqrt{{\lambda}/{2}}} dy \;
\end{eqnarray}
together with the well known inverse Laplace transform formula
\begin{eqnarray}
{LT}^{-1}_{\lambda \to x} e^{- y \sqrt{\lambda}} = \frac{y}{2 \sqrt{\pi} x^{3/2}} e^{-\frac{y^2}{4x}} \;.
\end{eqnarray}
One finally obtains an explicit formula for $P(x)$ 
\begin{equation}\label{exact_F}
P(x) = 4\big[ \sqrt{\frac{2}{\pi}} (1+2x^2) -  e^{2x^2} x (4x^2+3) {\rm erfc}(\sqrt{2}x)\big] \,,
\end{equation}
as announced in the introduction (\ref{P_scaling}) -- we recall that ${\rm erfc}(z) = (2/\sqrt{\pi})\int_z^\infty e^{-t^2} \, dt$ is the complementary error function. This PDF describes the typical fluctuations of the gaps (\ref{summary}), in particular, one can easily check that it is normalized, i.e. $\int_0^\infty P(x) dx = 1$. Its asymptotic behaviors are given in Eq. (\ref{asympt_P}).

\subsubsection{Large deviations of the gap.}\label{sec:largep0}

We now turn to the study of the large deviations of the gap~$d_{k,n}$ for large $n$ and large $k$, i.e. we study $\tilde p(z,\Delta)$ when $z \to 1$ but $\Delta = O(1)$. By expanding Eq. (\ref{expl_stationary}) for $z$ close to $1$ one finds
\begin{eqnarray}\label{large_p0_1}
\tilde p(z;\Delta) = 2 \left( \frac{\cosh(\Delta)}{\sinh^3(\Delta)} - \sqrt{1-z} \, \frac{2+\cosh(2 \Delta)}{\sinh^4(\Delta)}\right) + O((1-z)) \;,
\end{eqnarray}
from which it follows that for $k \gg 1$ one has
\begin{eqnarray}\label{phi_0}
p_k(\Delta) \sim \frac{1}{k^{3/2}}\varphi_0(\Delta) \;, \; \varphi_0(\Delta) = {\frac{1}{\sqrt{\pi}}} \frac{2 + \cosh{(2 \Delta)}}{\sinh^4(\Delta)}  \;.
\end{eqnarray}
Its asymptotic behaviors are given by
\bea\label{asympt_phi0}
\varphi_0(\Delta) \sim
\begin{cases}
&\dfrac{3}{\sqrt{\pi}} \, \Delta^{-4} = \dfrac{3\, \sigma_0^3}{\sqrt{8\pi}}\, \Delta^{-4} \;, \; \Delta \to 0 \\
& \\
& C_0\, e^{-2 \Delta} \;, \hspace*{2.cm} \; \Delta \to \infty \;,
\end{cases}
\eea
with $\sigma_0 = \sqrt{2}$ and $C_0 = 8/\sqrt{\pi}$. 

\section{The case of a jump PDF with $p=1$}\label{sec:p1}

In this subsection, we study the case where the jump PDF is given by Eq. (\ref{eq:2}) with $p=1$,~i.e., 
\begin{equation}\label{eq:jump1}
\begin{split}
f_1\left(\eta\right) = \frac{|\eta|}{2}e^{-{|\eta|}} \;.
\end{split}
\end{equation}
In this case, the standard deviation $\sigma_1$ is given by
\begin{eqnarray}
\sigma_1 = \sqrt{6} \;.
\eea
The goal is to obtain directly the GF of the PDF of the $k$-th gap $\tilde P(z,s;\Delta)$ (i.e. we do not consider the density of the $k$-th maximum $M_{k,n}$). This GF $\tilde P(z,s;\Delta)$ is given in terms of $\tilde Q(z,s;x,\Delta)$ and $\tilde R(z,s;x,\Delta)$ which are solutions of the coupled integral equations (\ref{eq:WH}). As discussed above, this type of Wiener-Hopf equations are extremely hard to solve. However, for the jump PDF in Eq. (\ref{eq:jump1}) these integral equations can be transformed into differential equations, as we did for the symmetric exponential density using the relation in Eq. (\ref{sec_deriv_exp}). This can be done thanks to a generalization of the relation (\ref{sec_deriv_exp_2}) to the present case (\ref{eq:jump1}). It reads
\begin{equation}\label{eq:4}
\begin{split}
\left(1-D^2\right)^2f_1\left(\eta\right) = \left(1+D^2\right)\delta\left(\eta\right)
\end{split}
\end{equation}
where we recall that $D =  \frac{d}{dx}$. 
Using this relation (\ref{eq:4}), we can turn the integral equations (\ref{eq:WH}), with the jump PDF in Eq. (\ref{eq:jump1}), into differential equations:
\begin{equation}\label{diff_syst_p1}
\begin{split}
&\left(1-D^2\right)^2\tilde{Q} = 1 + s\left(1+D^2\right)\tilde{Q}\\
&\left(1-D^2\right)^2\tilde{R} = 1 + zs\left(1+D^2\right)\tilde{R} \;,
\end{split}
\end{equation}
where $D\equiv \frac{d}{dx}$. As before for the $p=0$ case in Eqs. (\ref{expl_qcal1}) and (\ref{expl_pcal1}), the solutions of these linear equations (\ref{diff_syst_p1}) are of the form 
\begin{equation}\label{eq:5}
\begin{split}
&\tilde{Q}(z,s;x,\Delta) = A_1(z,s;\Delta)e^{-t_1(s){x}} + A_2(z,s;\Delta)e^{-t_2(s){x}} + \frac{1}{1-s}\\
&\tilde{R}(z,s;x,\Delta) = B_1(z,s;\Delta)e^{-t_1(zs){x}} + B_2(z,s;\Delta)e^{-t_2(zs){x}} + \frac{1}{1-zs}
\end{split}
\end{equation}
where $0<t_1(s)<t_2(s)$ are the two positive roots of the characteristic equation associated to the differential equation in (\ref{diff_syst_p1})
\begin{equation}\label{eq:roots_eq_p1}
\begin{split}
\left(1-t^2\right)^2 - s\left(1+t^2\right)=0 \;.
\end{split}
\end{equation}
Note that the two other negative roots $-t_1(s)$ and $-t_2(s)$ of the characteristic equation (\ref{eq:roots_eq_p1}) yield solutions for $\tilde Q$ and $\tilde R$ which are exponentially diverging at $x \to \infty$ and are thus not physically acceptable solutions. Such terms corresponding to these negative roots are thus not present in Eq.~(\ref{eq:5}). The roots $t_1(s)$ and $t_2(s)$ can still be computed explicitly for $p=1$, with the result
\begin{equation}\label{eq:roots}
\begin{split}
t_{1,2}(s) = \sqrt{\frac{s+2\mp\sqrt{s\left(s+8\right)}}{2}} \;. 
\end{split}
\end{equation}
In Fig. \ref{fig:roots_p1} we show the location of these roots(\ref{eq:roots}) in the complex plane as $s$ is varied from $s=0$ (in Fig. \ref{fig:roots_p1} a)) to $s=1$ (in Fig. \ref{fig:roots_p1} b)). An important feature to notice on this figure is that $t_1(s) \to 0$ as $s\to 1$ while $t_2(s) \to t_2(1) =t_2^* > 0$ in that limit. Indeed, in the limit $s \to 1$, one finds from Eq. (\ref{eq:roots}) that
\begin{eqnarray}\label{roots_t1_p1}
t_1(s) \sim c_1 \sqrt{1-s} \;\;\;\;{\rm with} \;\;\; c_1 = \frac{\sqrt{3}}{3} = \frac{\sqrt{2}}{\sigma_1}\;, \;\;\;{\rm as\;\;\;}s \to 1 \;,
\end{eqnarray}
while
\begin{eqnarray}\label{roots_t2_p1}
 \;\;\;\; t_2(s) \to t_2^* = \sqrt{3} \;, \;\;\;{\rm as\;\;\;}s \to 1 \;.
\end{eqnarray}

\begin{figure}[ht]
\centering
\begin{subfigure}{.45\textwidth}
  \centering
  \includegraphics[width=1.\linewidth]{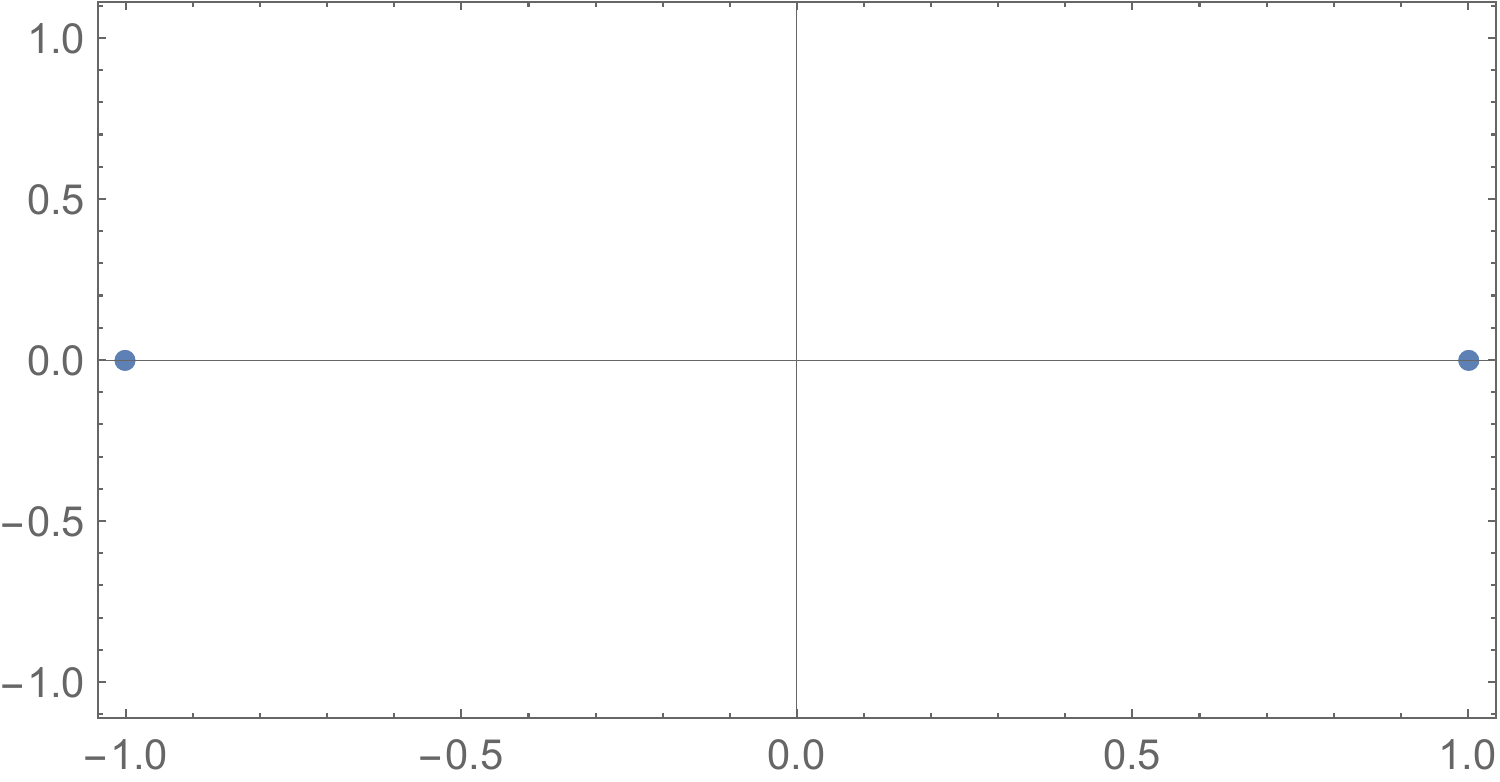}
  \caption{$s=0$}
\end{subfigure}%
\hspace{10mm}
\begin{subfigure}{.45\textwidth}
  \centering
  \includegraphics[width=1\linewidth]{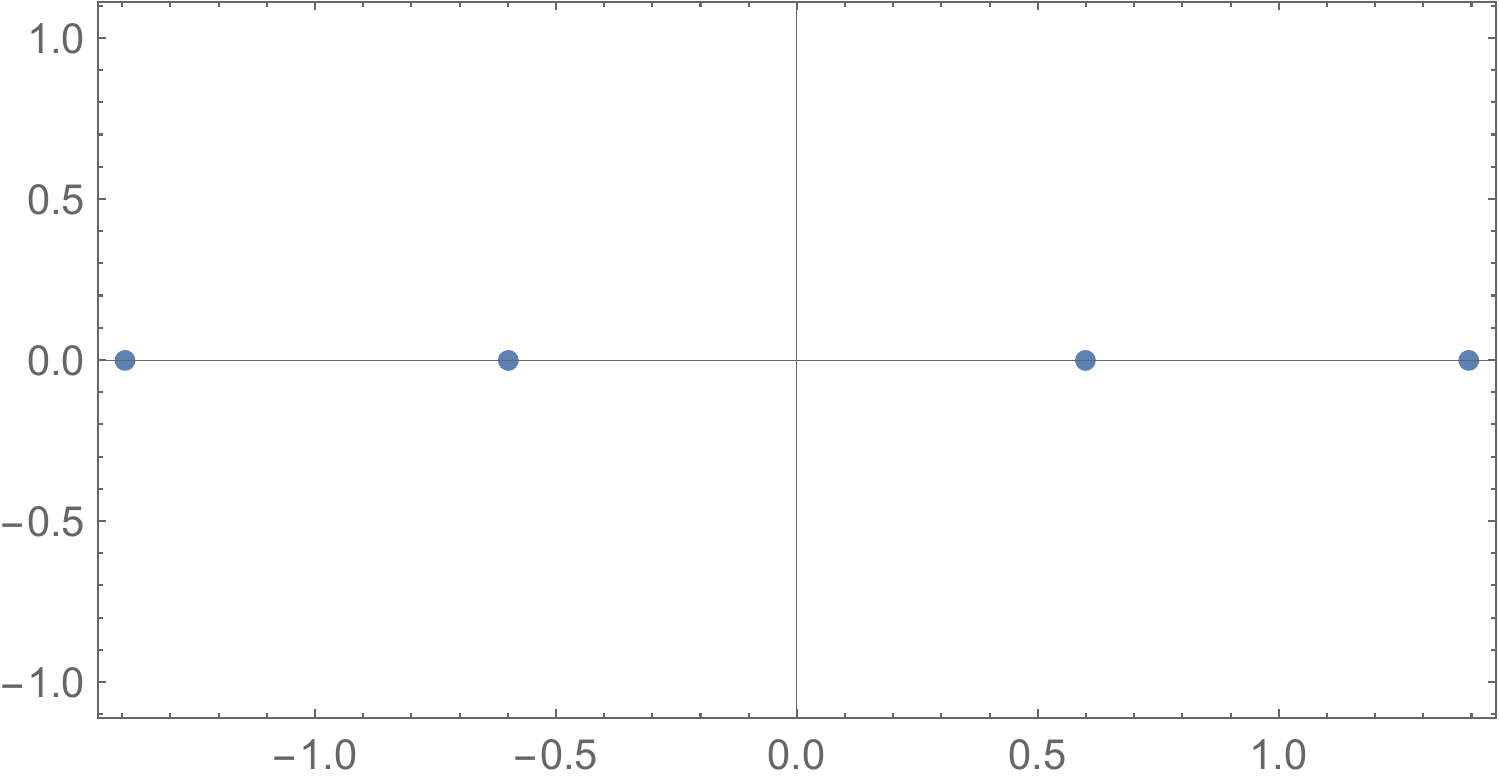}
  \caption{$s=0.3$}
\end{subfigure}
\begin{subfigure}{.45\textwidth}
  \centering
  \includegraphics[width=1\linewidth]{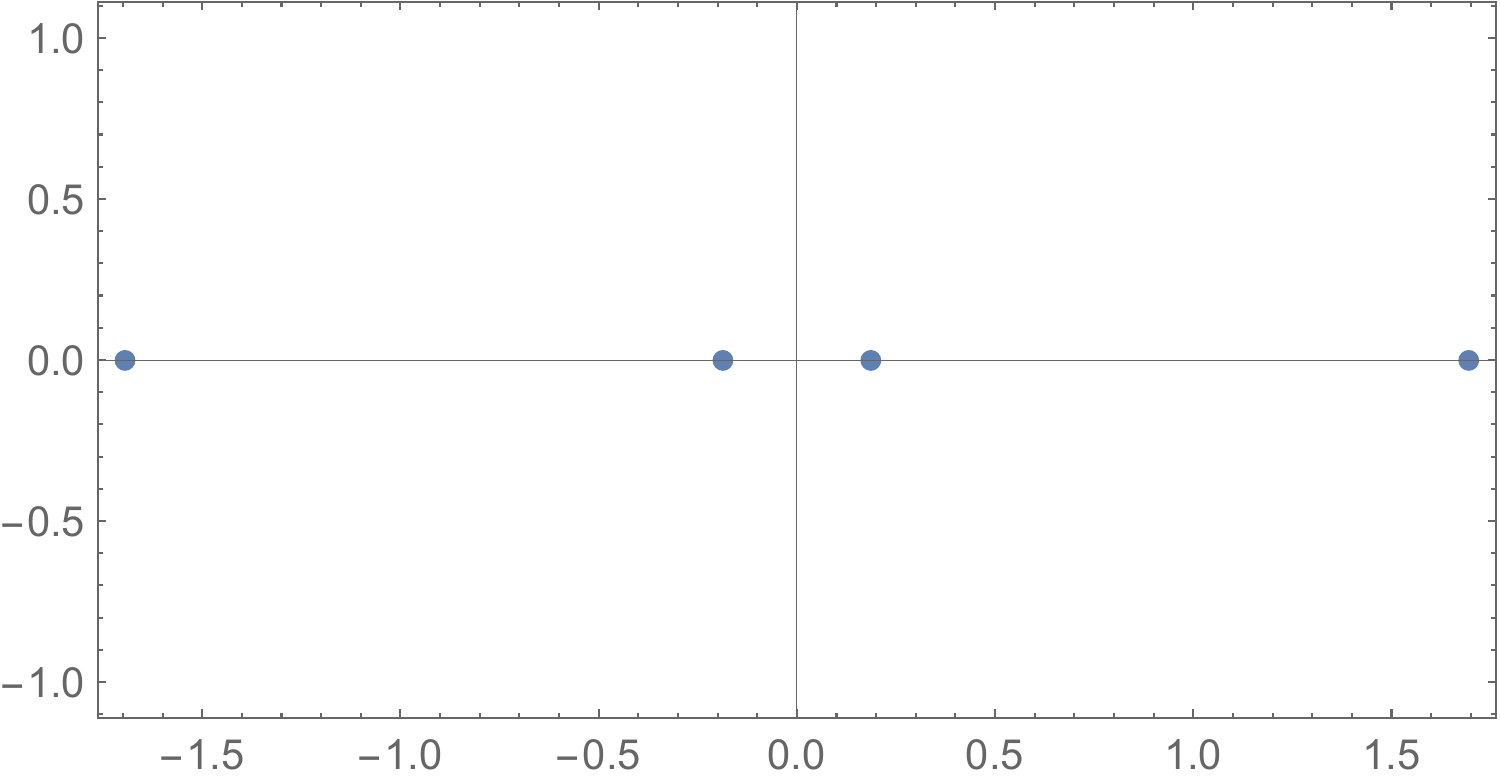}
  \caption{$s=0.9$}
\end{subfigure}%
\hspace*{10mm}
\begin{subfigure}{.45\textwidth}
  \centering
  \includegraphics[width=1\linewidth]{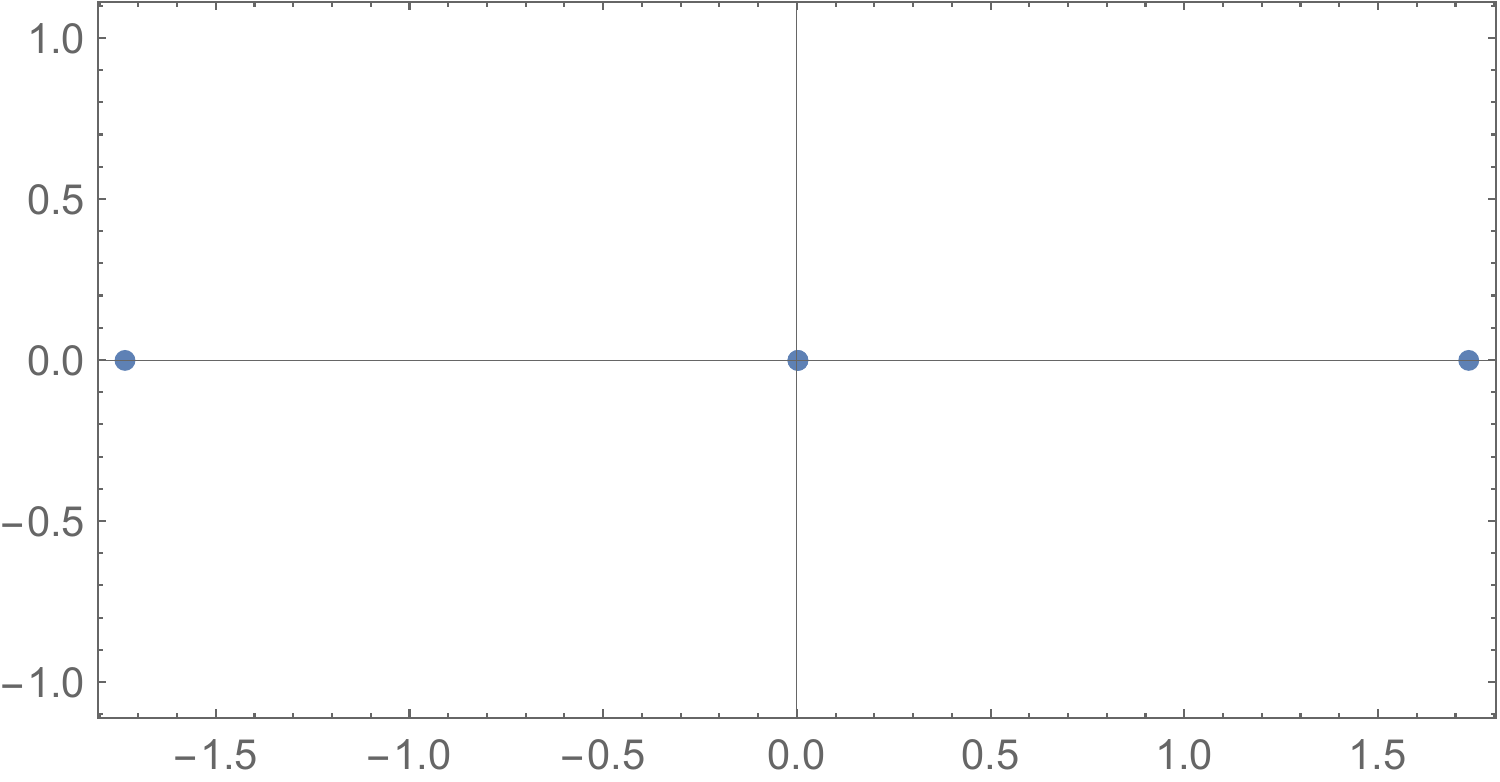}
  \caption{$s=1$}
\end{subfigure}
\caption{In these four figures we plot the 4 roots of (\ref{eq:roots_eq_p1}) for different values of $s$ in the complex plane (in this case all the roots are real). Note that for our purpose, only the {\it positive} roots $t_1(s)$ and $t_2(s)$ given in Eq. (\ref{eq:roots}) are relevant. For $s=0$ we have two solutions $\pm1$, each with double degeneracy. Then increasing $s$ we see that the degeneracy is lifted, giving rise to four roots. The two roots with the smallest modulus converge to $0$ as $s \to 1$ (in particular $t_1(s) \sim c_1\sqrt{1-s}$, see Eq. (\ref{roots_t1_p1})). The other two roots instead get farther and farther from the origin and converge to a finite value $\pm t_2^*$ as $s \to 1$.}
\label{fig:roots_p1}
\end{figure}
To compute the amplitudes $A_\alpha$'s and $B_\alpha$'s with $\alpha=1,2$ in Eqs. (\ref{eq:5}) we inject these expressions for $\tilde Q$ and $\tilde R$ in the original integral equations (\ref{eq:WH}). One obtains that they are given by the solution of the following $4 \times 4$ linear system of equations 
\begin{equation}\label{eq:ss}
\left\lbrace
\begin{split}
&\frac{A_1}{1-t_1} + \frac{A_2}{1-t_2}  - z\,e^{-\Delta}\left[\frac{B_1}{1+u_1} +\frac{B_2}{1+u_2}\right] + \frac{1}{1-s} - \frac{z\,e^{-\Delta}}{1-zs} = 0\\
&\frac{B_1}{1-u_1} + \frac{B_2}{1-u_2} - \frac{e^{-\Delta}}{z}\left[\frac{A_1}{1+t_1} + \frac{A_2}{1+t_2}\right] + \frac{1}{1-zs} - \frac{1}{z}\frac{e^{-\Delta}}{1-s} = 0\\
&\frac{A_1}{(1-t_1)^2} + \frac{A_2}{(1-t_2)^2} - z e^{-\Delta} \left[\frac{I_1(\Delta(1+u_1))}{(1+u_1)^2} B_1  +\frac{I_1(\Delta(1+u_2))}{(1+u_2)^2}\,B_2  \right] \\
&+ \frac{1}{1-s} - \frac{ze^{-\Delta}}{1-zs}I_1(\Delta) = 0\\
&\frac{B_1}{(1-u_1)^2} + \frac{B_2}{(1-u_2)^2} - \frac{e^{-\Delta}}{z} \left[\frac{I_1(\Delta(1+t_1))}{(1+t_1)^2}A_1  +  \frac{I_1(\Delta(1+t_2))}{(1+t_2)^2}\,A_2 \right] \\
&+ \frac{1}{1-zs} - \frac{1}{z}\frac{e^{-\Delta}}{1-s}I_1(\Delta) = 0
\end{split}
\right. \;,
\end{equation}
where $A_\alpha \equiv A_\alpha(z,s;\Delta)$ and similarly $B_\alpha \equiv B_\alpha(z,s;\Delta)$. In Eq. (\ref{eq:ss}) we also used $t_\alpha \equiv t_\alpha(s)$ as well as $u_\alpha \equiv t_{\alpha}(zs)$ and introduced the function $I_m(x)$ given~by
\begin{eqnarray}\label{def_In}
I_m(x) = \sum_{l=0}^m \frac{x^l}{l!} \;.
\end{eqnarray}
These amplitudes $A_\alpha$ and $B_\alpha$ can thus be computed explicitly (using {\it Mathematica} for instance), and one eventually obtains the GF of the  $k$-th gap density from Eq. (\ref{eq:6})~(see~\ref{sec:app:64})
\begin{eqnarray}\label{GF_pdf_delta_p=1}
\hspace*{-2.5cm}\tilde {P}(z,s;\Delta) &=& \partial_{\Delta} \left[ {A}_1(z,s; \Delta) + {A}_2(z,s; \Delta) \right] + \frac{1}{t_1(s)} \partial_\Delta^2 {A}_1(z,s; \Delta) +  \frac{1}{t_2(s)} \partial_\Delta^2 {A}_2(z,s; \Delta) \nonumber\\
&+& z \,e^{t_1(zs)\Delta} \left[\partial_{\Delta} {B}_1(z,s; \Delta) + \frac{1}{t_1(zs)} \partial_\Delta^2 {B}_1(z,s; {\Delta})\right] \\ &+&z \,e^{t_2(zs)\Delta} \left[\partial_{\Delta} {B}_2(z,s; \Delta) + \frac{1}{t_2(zs)} \partial_\Delta^2 {B}_2(z,s; {\Delta})\right] \nonumber 
 \end{eqnarray}
which is a generalization to $p=1$ of the result obtained for $p=0$ in Eq. (\ref{GF_pdf_delta}).

In principle this linear system can be exactly solved using \textit{Mathematica}. However, since we are mostly interested in the large $n$ limit of the PDF of the gap, we analyse this linear system in the limit $s \to 1$. Guided by the exact solution of the exponential case (\ref{asympt_s1_A}) and (\ref{asympt_s1_B}), we anticipate that in the limit $s \to 1$, the leading behaviors of $A_\alpha$'s and $B_\alpha$'s are given by 
%
%
%
%
\begin{equation}\label{eq:exp_ampl}
\begin{split}
&A_1(z,s;\Delta) \sim -\frac{1}{1-s} + \frac{A_{1,1}(z;\Delta)}{\sqrt{1-s}} \;,\hspace{8mm} A_2(z,s;\Delta)\sim \frac{A_{2,1}(z;\Delta)}{\sqrt{1-s}}\\
&B_1(z,s;\Delta) \sim  \frac{B_{1,1}(z;\Delta)}{\sqrt{1-s}}\;, \hspace{23mm} B_2(z,s;\Delta) \sim \frac{B_{2,1}(z;\Delta)}{\sqrt{1-s}} \;.
\end{split}
\end{equation}
Hence, by injecting this expansion (\ref{eq:exp_ampl}) in the linear system (\ref{eq:ss}) and using the behavior of the roots obtained in Eqs. (\ref{roots_t1_p1}) and (\ref{roots_t2_p1}) one obtains a linear system for $A_{\alpha,1}$ and $B_{\alpha,1}$ 
\begin{equation}\label{eq:ss_sto1}
\left\lbrace
\begin{split}
&A_{1,1} - c_1 + \frac{A_{2,1}}{1-t_2^*}  - z\,e^{-\Delta}\left[\frac{B_{1,1}}{1+t_1(z)} +\frac{B_{2,1}}{1+t_2(z)}\right]  = 0\\
&\frac{B_{1,1}}{1-t_1(z)} + \frac{B_{2,1}}{1-t_2(z)} - \frac{e^{-\Delta}}{z}\left[A_{1,1} + c_1 + \frac{A_{2,1}}{1+t_2^*}\right] = 0\\
&A_{1,1}- 2 c_1+ \frac{A_{2,1}}{(1-t^*_2)^2} - z \,e^{-\Delta} \left[\frac{I_1(\Delta(1+t_1(z)))}{(1+t_1(z))^2}\,B_{1,1} +\frac{I_1(\Delta(1+t_2(z)))}{(1+t_2(z))^2}\, B_{2,1}  \right] = 0\\
&\frac{B_{1,1}}{(1-t_1(z))^2} + \frac{B_{2,1}}{(1-t_2(z))^2} \\
&- \frac{e^{-\Delta}}{z} \left[I_1(\Delta)\,A_{1,1}+ c_1(2\,I_1(\Delta) - \Delta I_1'(\Delta))  + \frac{I_1(\Delta(1+t_2^*)) }{(1+t_2^*)^2} A_{2,1}\right] = 0 \\
\end{split}
\right. \;,
\end{equation}
where we recall that $c_1 = \sqrt{2}/\sigma_1$. One can check, using {\it Mathematica}, that this linear system (\ref{eq:ss_sto1}) admits a unique solution, although its explicit expression is of course a bit cumbersome. At this stage, from this expansion (\ref{eq:exp_ampl}) together with the asymptotic behavior of the roots in Eqs. (\ref{roots_t1_p1}) and (\ref{roots_t2_p1}), one can show, from Eq. (\ref{GF_pdf_delta_p=1}) that $P_{k,n}(\Delta)$ reaches a stationary limit when $n \to \infty$. Indeed, one observes that in the limit $s \to 1$ the leading term in the GF $\tilde P(z,s;\Delta)$ is the second term in Eq. (\ref{GF_pdf_delta_p=1}), i.e. the term proportional to $1/t_1(s)$ since $t_1(s) \sim c_1 \sqrt{1-s}$  as $s \to 1$ while $t_2^*$ is finite (\ref{roots_t2_p1}). Therefore, as in the case $p=0$ [see Eq. (\ref{stationary})] one has
\beq\label{stationary_p1}
\lim_{n \to \infty} P_{k,n}(\Delta) = p_k(\Delta)
\eeq
where the GF of $p_k(\Delta)$ is given by
\beq\label{stationary_p12}
\tilde p(z;\Delta) = \sum_{k=1}^\infty z^k \, p_k(\Delta) = \frac{1}{c_1} \partial^2_\Delta A_{1,1}(z;\Delta) \;,
\eeq
where $A_{1,1}(z;\Delta)$ is obtained by solving the linear system in Eq. (\ref{eq:ss_sto1}). We are mainly interested in the large $k$ limit of $p_k(\Delta)$, which can be obtained from the behavior of its GF $\tilde p(z;\Delta)$ in the limit $z \to 1$. We thus analyse the solution of this linear system (\ref{eq:ss_sto1}) in the limit $z\to 1$ and consider, as before, (i) the typical fluctuations (where $\Delta  = O(1/\sqrt{k})$) and (ii) the large deviations when $\Delta = O(1)$. 

\subsection{Typical fluctuations}\label{sec:typ_p1}

As in the case $p=0$, the regime of typical fluctuations correspond, for the GF $\tilde p(z;\Delta)$ to the limit $z \to 1$, $\Delta \to 0$, but keeping the scaling variable $\lambda = \sigma_1^2(1-z)/\Delta^2$ fixed [see Eq. (\ref{scaling_var})], where here $\sigma_1 = \sqrt{6}$. Inspired by the solution for $p=0$ [see Eq. (\ref{scaling_form_p0})], we look for a solution of the linear system (\ref{eq:ss_sto1}) of the form 
\bea\label{scal_typ_p1}
&A_{1,1}  \sim \frac{1}{\sqrt{1-z}} a_1(\lambda) + O(\sqrt{1-z}) \;, \; &B_{1,1}  \sim \frac{1}{\sqrt{1-z}} b_1(\lambda) + O(\sqrt{1-z}) \\
&A_{2,1} \sim a_2(\lambda) +   O(\sqrt{1-z}) \;, \; &B_{2,1} \sim b_2(\lambda) +  O(\sqrt{1-z})  \;.
\eea
Injecting these expansions (\ref{scal_typ_p1}) in the linear system in Eq. (\ref{eq:ss_sto1}) and expanding carefully the coefficients in limit $z \to 1$, $\Delta \to 0$, but keeping the scaling variable $\lambda = \sigma_1^2(1-z)/\Delta^2$ fixed, one finds that the functions $a_\alpha$'s satisfy a $2 \times 2$ linear system. To show how to get this linear system, let us consider, as an example, the left hand side of the first line  of Eq. (\ref{eq:ss_sto1}) in that limit 
\bea\label{example}
\hspace*{-2.2cm}&A_{1,1} - c_1 + \frac{A_{2,1}}{1-t_2^*}  - z\,e^{-\Delta}\left[\frac{B_{1,1}}{1+t_1(z)} +\frac{B_{2,1}}{1+t_2(z)}\right] \\
\hspace*{-2.2cm}&= \frac{a_1(\lambda)}{\sqrt{1-z}} - c_1 + \frac{a_2(\lambda)}{1-t_2^*} - \frac{b_2(\lambda)}{1+t_2^*}  - \frac{b_1(\lambda)}{\sqrt{1-z}}\left(1 - \sigma \frac{\sqrt{1-z}}{\sqrt{\lambda}}\right)(1 - c_1 \sqrt{1-z})  + O(\sqrt{1-z}) \nonumber
\eea
where, in the first line of Eq. (\ref{example}), we have replaced $A_{\alpha, 1}$ and $B_{\alpha, 1}$ (for $\alpha = 1,2$) by their asymptotic expansions in Eq. (\ref{scal_typ_p1}), $t_1(z)$ by $t_1(z) \sim c_1 \sqrt{1-z}$ and $t_2(z) \sim t_2^*$ [see Eq. (\ref{roots_t2_p1})]. Furthermore, we have replaced $\Delta$ by $\Delta = \sigma_1 \sqrt{(1-z)/\lambda}$, and then expanded $e^{-\Delta} = e^{-\sigma_1 \sqrt{(1-z)/\lambda}} \sim 1 - \sigma_1 \sqrt{(1-z)/\lambda}$ as $z \to 1$, to leading order. Collecting terms in the second line of Eq. (\ref{example}) one obtains
\bea\label{example2}
&\hspace*{-1.5cm}A_{1,1} - c_1 + \frac{A_{2,1}}{1-t_2^*}  - z\,e^{-\Delta}\left[\frac{B_{1,1}}{1+t_1(z)} +\frac{B_{2,1}}{1+t_2(z)}\right] \\
&\hspace*{-1.5cm}= (a_1(\lambda) - b_1(\lambda))\frac{1}{\sqrt{1-z}} - c_1 + \frac{a_2(\lambda)}{1-t_2^*} - \frac{b_2(\lambda)}{1+t_2^*}  + b_1(\lambda) \left(\frac{\sigma_1}{\sqrt{\lambda}} + c_1 \right) + O(\sqrt{1-z}) \;, \nonumber 
\eea
where $c_1 = \sqrt{2}/\sigma_1$. The three other equations of the linear system in Eq. (\ref{eq:ss_sto1}) can be analyzed along the same lines. By cancelling the terms of order $1/\sqrt{1-z}$, we immediately obtain [see for example Eq. (\ref{example2})]
\beq\label{aeqb}
a_1(\lambda) = b_1(\lambda) \;, \; a_2(\lambda) = b_2(\lambda) \;,
\eeq
while the cancellation of the constant term in these sets of equations leads to a $2\times2$ linear system for $a_\alpha$
\begin{equation}\label{linear_a_alpha1}
\left\lbrace
\begin{split}
& - c_1 + a_2(\lambda)\left(\frac{1}{1-t_2^*} - \frac{1}{1+t_2^*} \right) + a_1(\lambda) \left(\frac{\sigma_1}{\sqrt{\lambda}} + c_1 \right) = 0 \\
&-2 c_1 + a_2(\lambda)\left(\frac{1}{(1-t_2^*)^2} - \frac{1}{(1+t_2^*)^2}  \right) + 2\,c_1 a_1(\lambda)\ = 0
\end{split}
\right. \;.
\end{equation}
In fact, we only need to compute $a_1(\lambda)$ to get the GF of the PDF $p_k(\Delta)$ from  Eq. (\ref{stationary_p12}). It turns out that $a_1(\lambda)$ can be simply obtained by summing up the two lines of Eq. (\ref{linear_a_alpha1}). Indeed, by adding these two equalities we obtain
\begin{eqnarray}\label{linear_a_alpha2}
\hspace*{-1.5cm}a_1(\lambda)\left(3c_1 + \frac{\sigma_1}{\sqrt{\lambda}} \right) + a_2(\lambda)\left( \frac{1}{1-t_2^*} +\frac{1}{(1-t_2^*)^2}  -  \frac{1}{1+t_2^*} -  \frac{1}{(1+t_2^*)^2}\right) = 3 c_1 \;.
\end{eqnarray}
Using the identity
\bea\label{identity}
&\frac{1}{1-t} +\frac{1}{(1-t)^2}  -  \frac{1}{1+t} -  \frac{1}{(1+t)^2} \nonumber \\
&= \frac{2}{t(1+t)^2(1-t)^2} \left[1+t^2 - (1-t^2)^2 \right]
\eea
we see that the coefficient of $a_2(\lambda)$ in Eq. (\ref{linear_a_alpha2}) actually vanishes, since $t_2^*$ is solution of Eq. (\ref{eq:roots_eq_p1}) with $s=1$. Therefore, Eq. (\ref{linear_a_alpha2}) yields finally
\beq\label{a1_p1}
a_1(\lambda) = \frac{1}{1 + {\sigma_1}/{(3c_1 \sqrt{\lambda})}} = \frac{1}{1 + \sqrt{2/\lambda}} \;,
\eeq
where we have used $\sigma_1/(3c_1) = \sqrt{2}$ in this case. Using this result (\ref{a1_p1}) in Eq. (\ref{stationary_p12}) one obtains that, in this scaling limit, the GF $\tilde p(z;\Delta)$ takes the scaling form
\beq\label{typ_GF_p1}
\tilde p(z;\Delta) \sim \frac{\sigma_1^2}{\Delta^3} \frac{1}{(1+\sqrt{\lambda/2})^3} \;,
\eeq
which is exactly similar to the form found for the case $p=0$ in Eq. (\ref{scaling_form_p0bis}). Therefore, the gap density $p_k(\Delta)$ takes the same scaling form (\ref{scaling_form}), $p_k(\Delta) \sim \sqrt{k}/\sigma_1 P(\sqrt{k} \Delta/\sigma_1)$ with the same distribution $P(x)$ given in Eq. (\ref{exact_F}) and we recall that $\sigma_1 = \sqrt{6}$.

\subsection{Large fluctuations}\label{sec:large_devp1}

As in the case $p=0$ (see section \ref{sec:largep0}), the large deviations of the gap $d_{k,n}$ are obtained by studying the GF $\tilde p(z;\Delta)$ in the limit $z \to 1$ keeping $\Delta = O(1)$. The leading behavior of $\tilde p(z;\Delta)$ in this limit can be obtained from the analysis of the linear system (\ref{eq:ss_sto1}) using a similar approach as in section \ref{sec:typ_p1}. Indeed, we look for an expansion of the coefficients $A_{\alpha,1}$ and $B_{\alpha,1}$ in Eq. (\ref{eq:exp_ampl}) of the form 
\bea\label{exp_large_dev_p1}
\begin{split}
&A_{\alpha,1}(z;\Delta) = A_{\alpha,1}^{(0)}(\Delta) - A_{\alpha,1}^{(1)}(\Delta) \sqrt{1-z} + O(1-z) \\
&B_{\alpha,1}(z;\Delta) = B_{\alpha,1}^{(0)}(\Delta) - B_{\alpha,1}^{(1)}(\Delta) \sqrt{1-z} + O(1-z) 
\end{split}
\eea
for $\alpha = 1,2$ and inject these expansions in the linear system (\ref{eq:ss_sto1}). By doing this, one obtains a linear system for the functions $A_{\alpha,1}^{(0)}, A_{\alpha,1}^{(1)}$ and $B_{\alpha,1}^{(0)}, B_{\alpha,1}^{(1)}$ for $\alpha =1,2$. This expansion (\ref{exp_large_dev_p1}) together with Eq. (\ref{stationary_p12}) shows that 
in this limit $z \to 1$ keeping $\Delta = O(1)$, $\tilde p(z;\Delta)$ admits an expansion similar to the case $p=0$ given in Eq. (\ref{large_p0_1}), i.e. 
\begin{equation}
\begin{split}\label{large_p1}
\tilde{p}\left(z;\Delta\right) = \frac{1}{c_1} \partial_\Delta^2  A_{1,1}^{(0)}(\Delta) - \sqrt{1-z}\frac{1}{c_1} \partial_\Delta^2  A_{1,1}^{(1)}(\Delta) + O(1-z) \;.
\end{split}
\end{equation}
From Eq. (\ref{large_p0_1}) one immediately gets the large $k$ behavior of $p_k(\Delta)$ 
\beq\label{phi1_p1}
p_k(\Delta) \sim \frac{1}{k^{3/2}} \varphi_1(\Delta) \;, \;\;\; \varphi_1(\Delta) = \frac{1}{2 \sqrt{\pi}\,c_1} \partial_\Delta^2 A_{1,1}^{(1)}(\Delta) \;,
\eeq
where $c_1 = \sqrt{2}/\sigma_1$. The full expression of $\varphi_1(\Delta)$ is quite cumbersome but its asymptotic behaviors for $\Delta \to 0$ and $\Delta \to \infty$ can be obtained by analysing the linear system for the functions $A_{\alpha,1}^{(0)}, A_{\alpha,1}^{(1)}$ and $B_{\alpha,1}^{(0)}, B_{\alpha,1}^{(1)}$ for $\alpha =1,2$ in these limits. One obtains \cite{Matteo}

\begin{eqnarray}\label{asympt_phi1}
\begin{split}
\varphi_1(\Delta) \sim 
\begin{cases}
&9\,\sqrt{\dfrac{3}{\pi}}\,\Delta^{-4} = \dfrac{3\,\sigma_1^3}{\sqrt{8 \pi}}\,\Delta^{-4} \;, \; \Delta\to0\\
& \\
&C_1\, \Delta^2e^{-2\Delta} \;, \; \hspace*{1.85cm}\Delta \to\infty
\end{cases}
\end{split}
\end{eqnarray}
%
where we have used $\sigma_1 = \sqrt{6}$ and where the amplitude $C_1$ can be computed explicitly
\begin{equation}\label{C1}
\begin{split}
C_1=\frac{16}{\sqrt{\pi}}\frac{2702\sqrt{\pi}+1560}{18817+679\sqrt{3}}\approx2.86670\ldots
\end{split}
\end{equation}

In summary, in the case $p=1$, we have shown that the gap PDF $P_{k,n}(\Delta)$ reaches a limiting density $p_k(\Delta)$ as $n \to \infty$ [see Eq. (\ref{stationary_p1})].  In the limit of large $k$, it displays both a typical regime for $\Delta = O(k^{-1/2})$ and a large deviation regime for $\Delta = O(1)$, which is summarized as follows   
\begin{eqnarray}\label{summary_p1}
p_k(\Delta) \sim
\begin{cases}
 &\sqrt{k}/\sigma_1 P(\sqrt{k} \, \Delta/\sigma_1) \;, \; \Delta = O(k^{-1/2}) \\
 &\\
 &k^{-3/2} \varphi_1(\Delta) \;, \; \hspace*{0.8cm} \Delta = O(1) \;,
\end{cases}
\end{eqnarray}
where the scaling function $P(x)$ is the same as for $p=0$ (\ref{exact_F}) while the function $\varphi_1(\Delta)$ is different from $\varphi_0(\Delta)$ as it can be seen by comparing the asymptotic behaviors in Eq. (\ref{asympt_phi1}) and (\ref{asympt_phi0}) respectively. Despite of these differences, the small $\Delta$ behavior can be written in both cases as $\varphi_p(\Delta) \sim (3/\sqrt{8 \pi}) \sigma_p^3 \Delta^{-4}$ for $p=0,1$ such that this behavior matches with the right tail of the central part describing the typical fluctuations of $\Delta = O(k^{-1/2})$ (see the discussion around Eqs. (\ref{matching_1}) and (\ref{matching_2}) in Introduction).

%

\section{The case of arbitrary integer $p$}\label{sec:genp}

In this section, we show that the analysis carried out in section \ref{sec:p1} can be extended to the wider class of jump PDF
\beq\label{f_eta_p}
f_p(\eta) = \frac{|\eta|^p}{2\,p!} e^{-|\eta|} \;,
\eeq
with any arbitrary integer value $p \geq 0$, and for which the variance $\sigma^2_p$ is given by
\beq\label{var_p}
\sigma^2_p = {(p+1)(p+2)} \;.
\eeq
As before, the starting point of our analysis is the expression of the jump PDF $P_{k,n}(\Delta)$ in Eq.~(\ref{eq:kgap}) in terms of the auxiliary functions $Q$ and $R$ which satisfy the coupled recursion relations in Eqs.~(\ref{back_fptilde1}) and (\ref{back_fptilde2}). As we have seen, it is more convenient to work with the double GF $\tilde P(z,s;\Delta)$ which can be expressed in terms of the GF $\tilde Q$ and $\tilde R$ as in Eq. (\ref{eq:6}) which, in turn, satisfy \textit{Wiener-Hopf} integral equations given in (\ref{eq:WH}). The key-point is that these integral equations, for this choice of jump PDF (\ref{f_eta_p}), can be reduced to ordinary differential equations of order $(2p+2)$. This can be done using the following relation, which generalizes Eq. (\ref{eq:4}) to arbitrary integer $p$, namely
\begin{equation}\label{eq:3bis}
\begin{split}
\left(1-D^2\right)^{p+1}f_p(\eta) = \frac{1}{2}\left[\left(1-D\right)^{p+1}+\left(1+D\right)^{p+1}\right]\delta\left(\eta\right) \;,
\end{split}
\end{equation}
with $D = \frac{d}{dx}$. This identity (\ref{eq:3bis}) applied to the integral equations satisfied by $\tilde Q$ and $\tilde R$ in Eq. (\ref{eq:WH}) yields the following differential equations
\bea
&(1-D^2)^{p+1} \tilde Q = 1+ \frac{s}{2} \left( (1-D)^{p+1} + (1+D)^{p+1} \right) \tilde Q \label{ODEQ}\\
&(1-D^2)^{p+1} \tilde R = 1+ \frac{zs}{2} \left( (1-D)^{p+1} + (1+D)^{p+1} \right) \tilde R \;,\label{ODER}
\eea
valid for arbitrary integer $p \geq 0$. In particular for $p=1$ one recovers Eq. (\ref{diff_syst_p1}). As before [see Eq. (\ref{eq:5})] the solutions of these differential equations are obtained as superpositions of exponentials   
%
%
\begin{equation}\label{eq:7}
\begin{split}
&\tilde{Q}\left(z,s;x,\Delta\right) = \frac{1}{1-s} + \sum_{\alpha=1}^{p+1}A_\alpha(z,s;\Delta)e^{-t_\alpha(s)x}\\
   &\tilde{R}(z,s;x,\Delta) = \frac{1}{1-zs} + \sum_{\alpha=1}^{p+1}B_\alpha\left(z,s;\Delta\right)e^{-t_\alpha(zs){x}}
\end{split}
\end{equation}
where $t_\alpha(s)$ are the roots, with positive real parts ($\Re(t_\alpha(s)) > 0$), of the characteristic equation
\begin{equation}\label{eq:roots_eq}
\begin{split}
 \left(1-t^2\right)^{p+1} -\frac{s}{2}\left[\left(1+t\right)^{p+1} + \left(1-t\right)^{p+1}\right]=0
\end{split}
\end{equation}
and we order them such that $\Re(t_1)<\Re(t_2)<\cdots< \Re(t_{p+1})$. In Fig. \ref{fig:roots_pgen} we show the locations of these roots in the complex plane for $p=20$. In the large $p$ limit, they tend to cluster uniformly on the two circles of radius 1 and centered at $(-1,0)$ and $(+1,0)$. For our purpose, it is important to understand the behavior of these roots as $s \to 1$. In this limit, one can show (see Fig. \ref{fig:roots_pgen}) that
\begin{eqnarray}
&&\lim_{s\to 1} t_1(s) = 0 \;, \label{roots_t1} \\ 
&& \lim_{s \to 1} t_{\alpha}(s) = t_\alpha^* \neq 0 \;, \; \alpha \geq 2 \;. \label{roots_ta}
\end{eqnarray} 
By inspection of the characteristic equation (\ref{eq:roots_eq}) it is easy to see that $t_1(s)$ vanishes as as square root as $s \to 1$, i.e.
\begin{equation}\label{asympt_t1}
t_1(s) \sim c_p \sqrt{1-s} \;, {\rm with}\;\; c_p = \sqrt{\frac{2}{\left(p+1\right)\left(p+2\right)}} = \frac{\sqrt{2}}{\sigma_p} \;,
\end{equation}
where, in the last equality, we have used the expression of $\sigma_p$ given in Eq. (\ref{var_p}). It is interesting to notice that the mathematical structure of the problem studied here bears some similarities with a completely different problem related to the thermodynamics of a quantum Ising spin chain in presence of random magnetic fields, which are gamma distributed, like the jumps (\ref{eq:2}) in our problem \cite{luck}. In particular, exactly the same characteristic equation (\ref{eq:roots_eq}) for $s=1$ is studied in detail there (see Section 6.2 in \cite{luck}).

As before, to compute the amplitudes $A_\alpha$'s and $B_\alpha$'s with $\alpha = 1, \cdots, p+1$ in Eq. (\ref{eq:7}) we inject these expressions for $\tilde Q$ and $\tilde R$ in the original integral equations (\ref{eq:WH}). One thus obtains a linear system of $2(p+1)$ equations (for $2(p+1)$ unknown $A_\alpha$'s and $B_\alpha$'s). They read, for $r = 0, 1, \cdots, p$ (see \ref{app:integrals})
\begin{eqnarray}\label{eq:sss}
&&\hspace*{-2.2cm}\sum_{\alpha=1}^{p+1}\dfrac{A_\alpha}{\left(1-t_\alpha\right)^{r+1}} - z\,e^{-\Delta}\sum_{\alpha=1}^{p+1} \frac{B_\alpha}{(1+u_\alpha)^{r+1}} I_{r}(\Delta(1+u_\alpha)) +\frac{1}{1-s} - \frac{ze^{-\Delta}}{1-zs}I_{r}(\Delta) = 0 \nonumber \\
&&\hspace*{-2.2cm}\sum_{\alpha=1}^{p+1}\frac{B_\alpha}{\left(1-u_\alpha\right)^{r+1}} - \frac{e^{-\Delta}}{z}\sum_{\alpha=1}^{p+1} \frac{A_\alpha}{(1+t_\alpha)^{r+1}} I_{r}(\Delta(1+t_\alpha)) +\frac{1}{1-zs} -  \frac{e^{-\Delta}}{z(1-s)} I_{r}(\Delta) = 0 \nonumber \\ 
\end{eqnarray}
where we recall that $I_m(x) = \sum_{l=0}^m x^l/l!$ and we used the notation $u_\alpha = t_\alpha(zs)$. One can check that for $p=1$ this linear system (\ref{eq:sss}) for $r=0,1$ yields back (\ref{eq:ss}). In principle, the amplitudes $A_\alpha$ and $B_\alpha$ are obtained by solving this linear system of $2(p+1)$ equations. 
%
%
\begin{figure}[ht]
\centering
\begin{subfigure}{.45\textwidth}
  \centering
  \includegraphics[width=1.\linewidth]{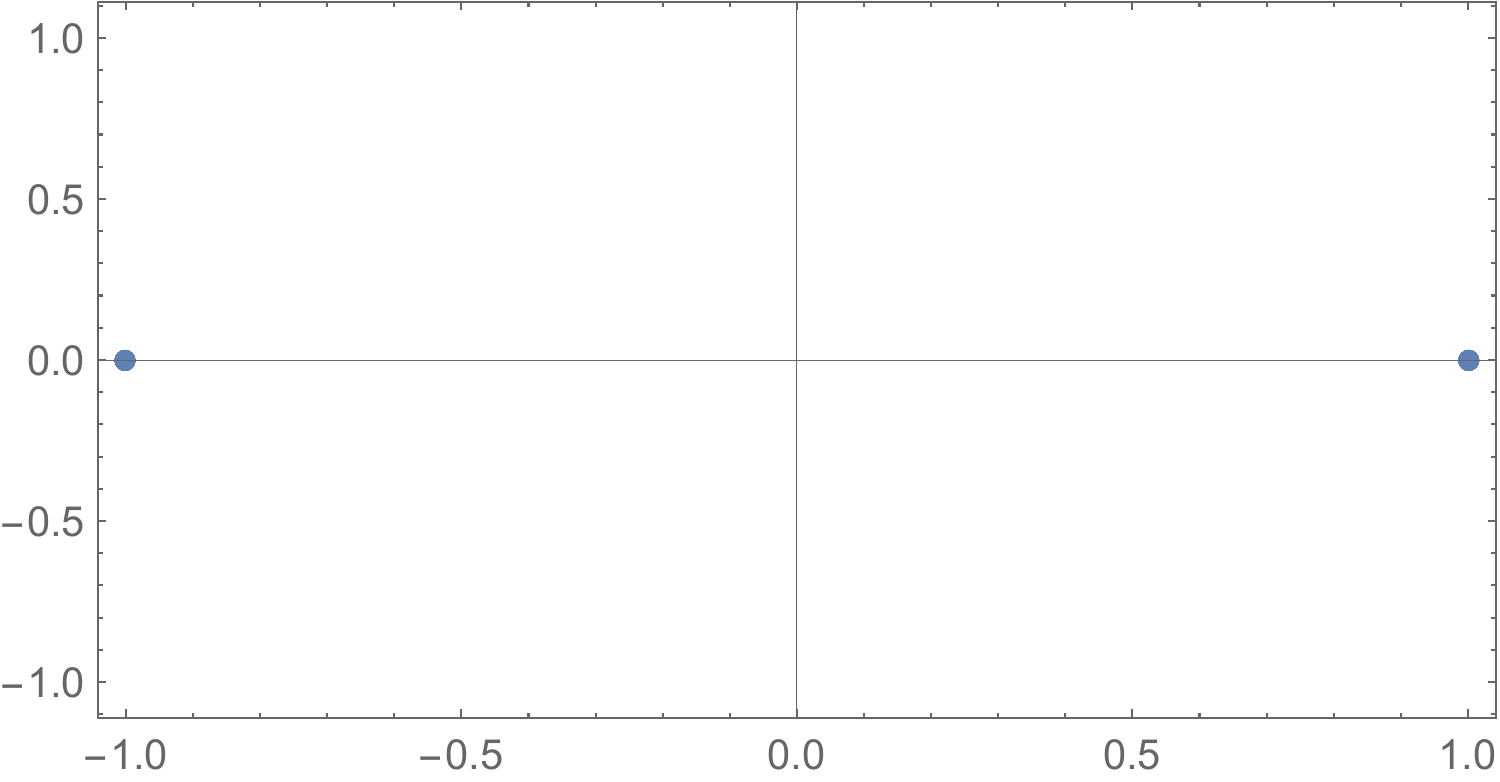}
  \caption{$s=0$}
\end{subfigure}%
\hspace{10mm}
\begin{subfigure}{.45\textwidth}
  \centering
  \includegraphics[width=1\linewidth]{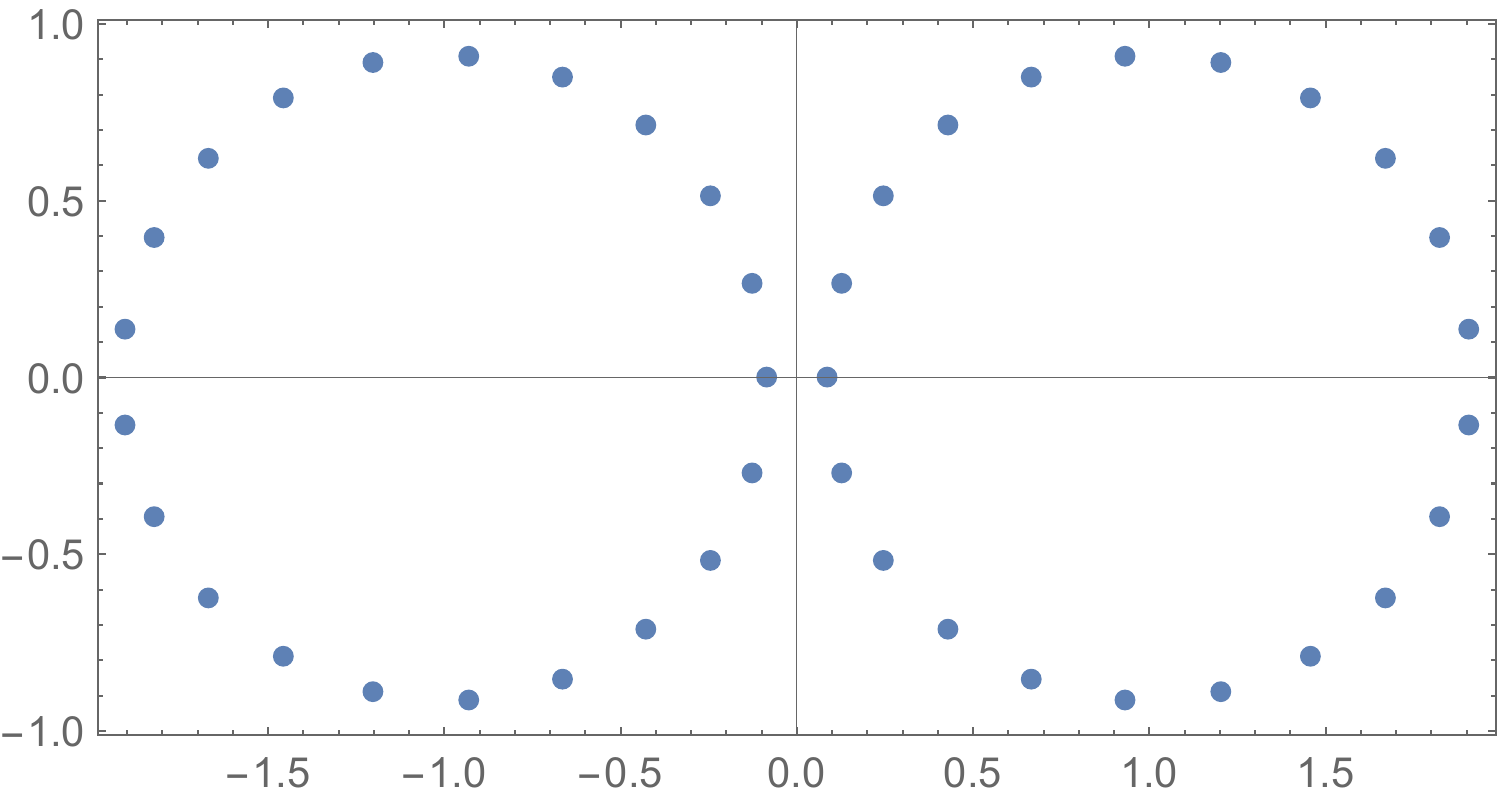}
  \caption{$s=0.3$}
\end{subfigure}
\begin{subfigure}{.45\textwidth}
  \centering
  \includegraphics[width=1\linewidth]{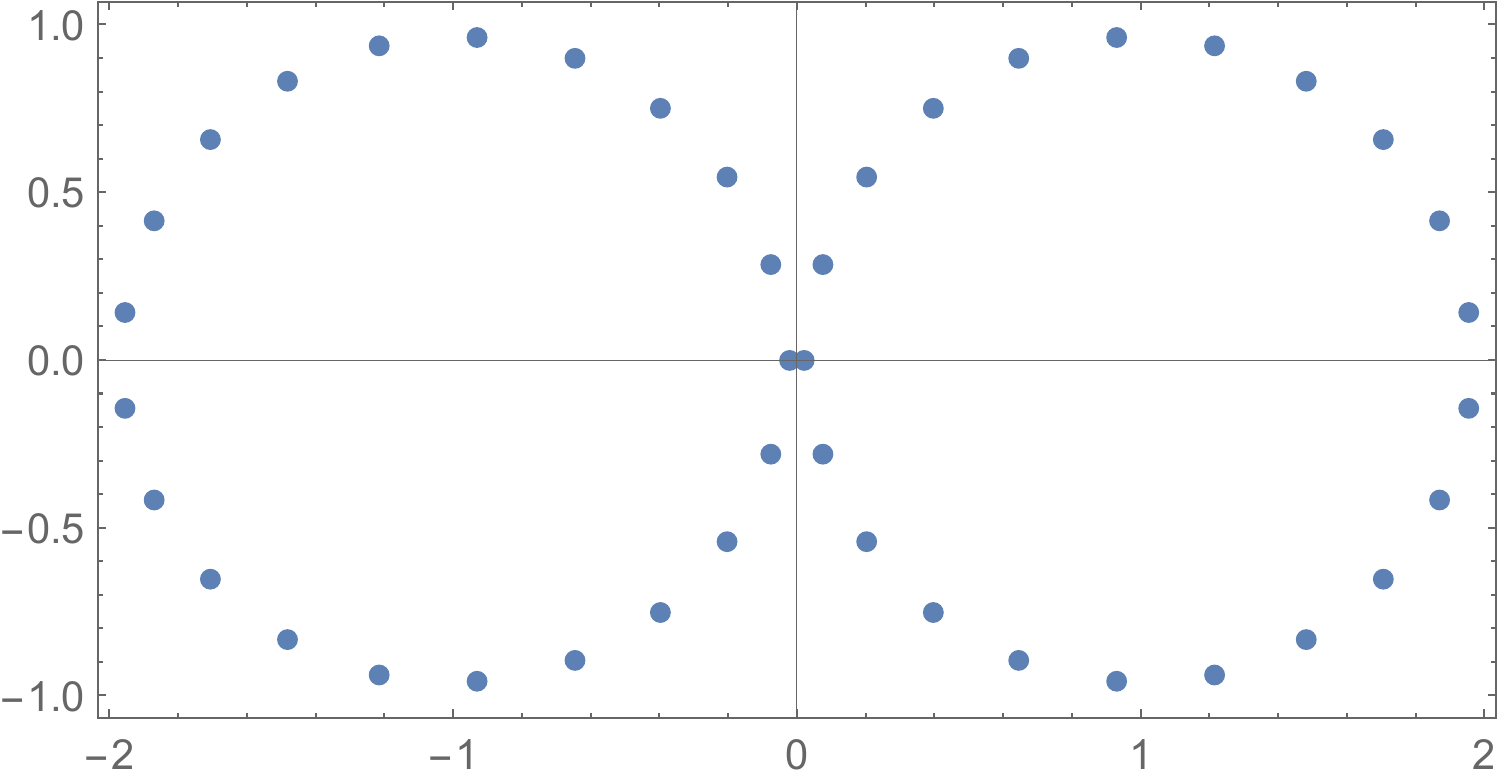}
  \caption{$s=0.9$}
\end{subfigure}%
\hspace*{10mm}
\begin{subfigure}{.45\textwidth}
  \centering
  \includegraphics[width=1\linewidth]{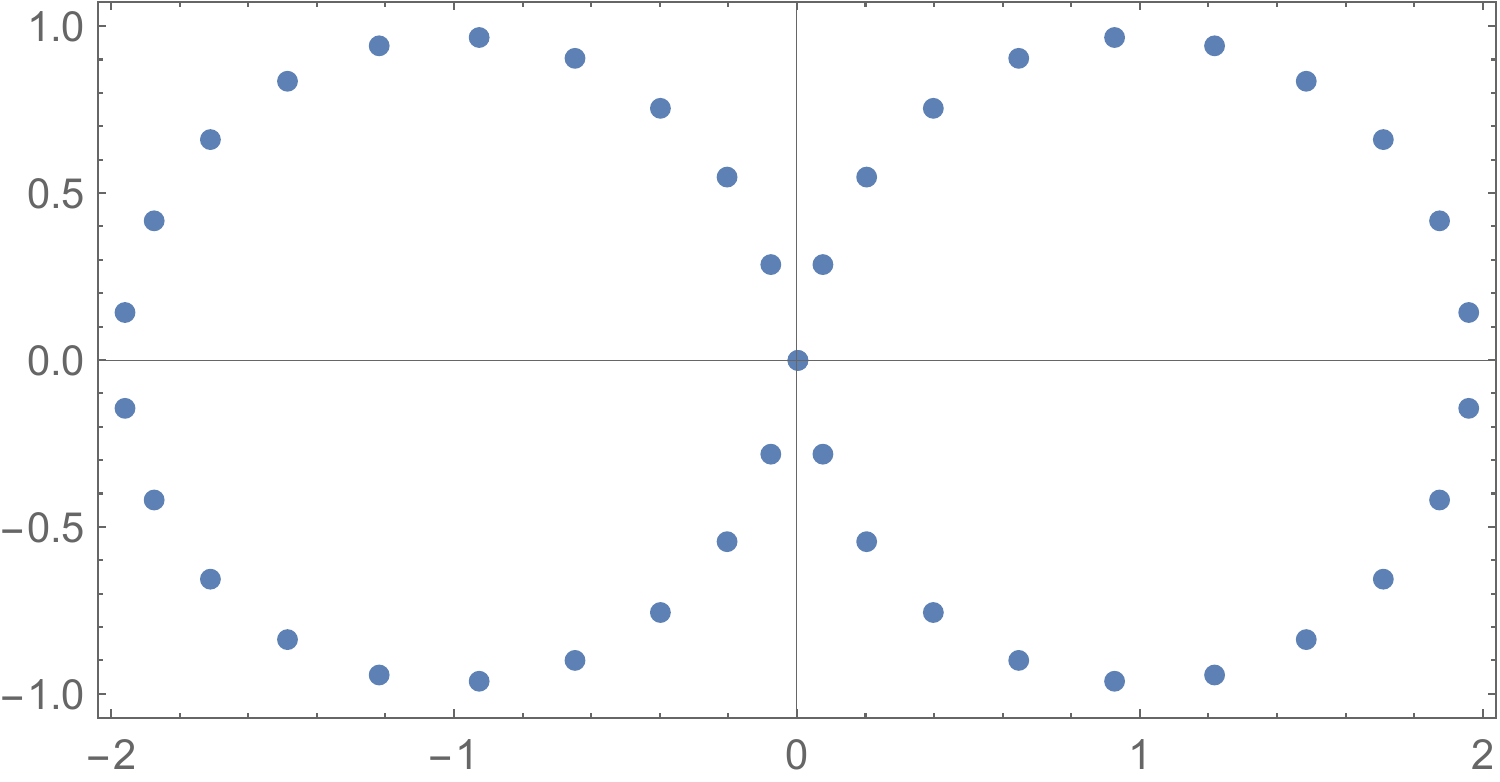}
  \caption{$s=1$}
\end{subfigure}
\caption{Location of the roots, in the complex plane, of Eq. (\ref{eq:roots_eq}) with $p=20$ and for different values of $s$. For $s=0$ there are two solutions $t=\pm1$, each with degeneracy \textit{twenty-one}. Then, as $s$ increases the degeneracy disappears giving rise to \textit{forty-two} distinct roots. The two roots with smallest magnitude converge to 0 as $s \to 1$ while the other forty roots remain finite in that limit.}
\label{fig:roots_pgen}
\end{figure}
From these amplitudes, the GF $\tilde P(z, s;\Delta)$ is then obtained by injecting (\ref{eq:7}) into (\ref{eq:6}) from which we get (see \ref{sec:app:64})
\begin{eqnarray}\label{eq:full_prob}
\hspace*{-2.5cm}\tilde{P}\left(z,s;\Delta\right) = \sum_{\alpha=1}^{p+1}\left[\left(\partial_\Delta+\frac{1}{t_\alpha(s)}\partial_\Delta^2\right)A_\alpha\left(z,s;\Delta\right) + z\,e^{t_\alpha(zs)}\left(\partial_\Delta+\frac{1}{t_\alpha(zs)}\partial^2_\Delta\right)B_\alpha\left(z,s;\Delta\right)\right].
\end{eqnarray}


Since we are interested in the large $n$ behavior of $P_{k,n}(\Delta)$, we now analyse the solution of this linear system in the limit $s \to 1$. In this limit, inspired by the solution found for $p=0$ [Eqs. (\ref{asympt_s1_A}) and (\ref{asympt_s1_B})] and $p=1$ (\ref{eq:exp_ampl}), we search for a solution under the form
\begin{eqnarray}\label{asympt_genp}
&&A_{1}(z,s;\Delta) \sim -\frac{1}{1-s} + \frac{A_{1,1}(z;\Delta)}{\sqrt{1-s}} \;, \; A_{\alpha}(z,s;\Delta) \sim \frac{A_{\alpha,1}(z;\Delta)}{\sqrt{1-s}} \;, \; \alpha \geq 2 \;, \\
&& B_{1}(z,s;\Delta) \sim \frac{B_{1,1}(z;\Delta)}{\sqrt{1-s}} \;, \; B_{\alpha}(z,s;\Delta) \sim \frac{B_{\alpha,1}(z;\Delta)}{\sqrt{1-s}} \;, \; \alpha \geq 2 
\end{eqnarray}
By injecting these forms (\ref{asympt_genp}) in the linear system (\ref{eq:sss}) and using the behavior of the roots as $s\to 1$ (\ref{roots_t1})-(\ref{asympt_t1}), one obtains straightforwardly a linear system of equations for the amplitudes $A_{\alpha,1}$ and $B_{\alpha,1}$ for $\alpha = 1,\cdots, p+1$. It reads, for $r = 0,1, \cdots, p$
\begin{eqnarray}\label{eq:sssto1}
&&\hspace*{-2.2cm} A_{1,1} + \sum_{\alpha=2}^{p+1}\dfrac{A_{\alpha,1}}{\left(1-t_\alpha^*\right)^{r+1}} - z\,e^{-\Delta}\sum_{\alpha=1}^{p+1} \frac{B_{\alpha,1}}{(1+t_\alpha(z))^{r+1}} I_{r}(\Delta(1+t_\alpha(z))) = c_p(r+1) \nonumber \\
&&\hspace*{-2.2cm}\sum_{\alpha=1}^{p+1}\frac{B_{\alpha,1}}{\left(1-t_\alpha(z)\right)^{r+1}} - \frac{e^{-\Delta}}{z}\left( A_{1,1} I_{r}(\Delta) + \sum_{\alpha=2}^{p+1} \frac{A_{\alpha,1}}{(1+t^*_\alpha)^{r+1}} I_{r}(\Delta(1+t_\alpha^*))  \right) \\
&&= c_p\,\frac{e^{-\Delta}}{z} \left((r+1)I_{r}(\Delta) - \Delta\,I'_{r}(\Delta) \right) \;. \nonumber \\ 
\end{eqnarray}
Solving explicitly this linear system is still very cumbersome. However, as in the previous cases $p=0$ and $p=1$, one can already show that the PDF of the $k$-th gap $P_{k,n}(\Delta)$ reaches a stationary density as $n \to \infty$ (provided the solution of this linear system (\ref{eq:sssto1}) exists of course). Indeed, by injecting the asymptotic behaviors of the amplitudes (\ref{asympt_genp})  and of the roots [(\ref{roots_t1})-(\ref{asympt_t1})] in the formula for the GF of $P_{k,n}(\Delta)$ one obtains by the same reasoning as in the case $p=1$ (see Eq. (\ref{eq:ss_sto1}) and below)
\begin{eqnarray}\label{stat_genp}
P_{k,n}(\Delta) \to p_k(\Delta) \;,
\end{eqnarray} 
where the GF of $p_k(\Delta)$ is given by
\begin{eqnarray}\label{stat_genp_GF}
\tilde p(z;\Delta) = \sum_{k=0}^\infty p_k(\Delta) \,z^k = \frac{1}{c_p} \partial^2_{\Delta} A_{1,1}(z;\Delta) \;,
\end{eqnarray}
where $A_{1,1}(z;\Delta)$ is given by the solution of the linear system (\ref{eq:sssto1}). 

As we did before, we now extract the leading behavior of $A_{1,1}(z;\Delta)$ as $z \to 1$, to obtain the 
large $k$ limit of $p_{k}(\Delta)$. Here also, we treat separately (i) the typical regime where $\Delta = O(1/\sqrt{k})$ and (ii) the large fluctuations regime where $\Delta = O(1)$.

\subsection{Typical fluctuations}

As we have seen before, the regime of typical fluctuations corresponds, for the GF $\tilde p(z;\Delta)$, to the limit $z \to 1$, $\Delta \to 0$, but keeping the scaling variable $\lambda = \sigma_p^2(1-z)/\Delta^2$ fixed [see Eq. (\ref{scaling_var})], where here $\sigma_p = \sqrt{(p+1)(p+2)}$. Guided by the solution found for $p=0$ (\ref{scaling_form_p0}) and $p=1$ (\ref{scal_typ_p1}) we look for a solution of the form
\bea\label{expansion_zto1_p}
&A_{1,1} \sim \frac{a_1(\lambda)}{\sqrt{1-z}} + O(\sqrt{1-z}) \;, \; &B_{1,1} \sim \frac{b_1(\lambda)}{\sqrt{1-z}} \;, \\
&A_{\alpha,1} \sim a_\alpha(\lambda) + O(\sqrt{1-z}) \;, \; &B_{\alpha,1} \sim {b_\alpha(\lambda)} + O(\sqrt{1-z}) \;, \, {\rm for} \;\; \alpha \geq 2 \;.
\eea
Performing the asymptotic analysis of the linear system (\ref{eq:sssto1}) as explained in Eq. (\ref{example}) for the case $p=1$, one first obtains that
\beq
a_{\alpha}(\lambda) = b_\alpha(\lambda) \;, \; \forall \alpha = 0, 1, \cdots, p
\eeq
while the $a_\alpha$'s are determined by the following linear system
\beq
a_1\left[(r+1)c_p + \frac{\sigma_p}{\sqrt{\lambda}}(1-I_r'(0))\right] + \sum_{\alpha=2}^{p+1} a_\alpha \left[\frac{1}{(1-t_\alpha^*)^{r+1}} - \frac{1}{(1+t_\alpha^*)^{r+1}} \right] = c_p(r+1)
\eeq
with $r = 0,1,\cdots, p$ and where $c_p$ is given in Eq. (\ref{asympt_t1}). Noting that $1-I'_r(0) = \delta_{r,0}$ one obtains
\beq\label{final_system}
a_1\left[(r+1)c_p + \frac{\sigma_p}{\sqrt{\lambda}}\delta_{r,0}\right] + \sum_{\alpha=2}^{p+1} a_\alpha \left[\frac{1}{(1-t_\alpha^*)^{r+1}} - \frac{1}{(1+t_\alpha^*)^{r+1}} \right] = c_p(r+1) \;,
\eeq
for $r=  0,1,\cdots,p$. In the case $p=1$, this set of equations (\ref{final_system}) yields back the system in Eq. (\ref{linear_a_alpha1}). Indeed, the case $r=0$ corresponds to the first relation in Eq. (\ref{linear_a_alpha1}) while the case $r=1$ corresponds to the second line of Eq. (\ref{linear_a_alpha1}). From Eq. (\ref{stat_genp_GF}), we see that we only need to know $a_1(\lambda)$ to compute $\tilde p(z;\Delta)$. Remarkably, we can easily obtain $a_1(\lambda)$ from (\ref{final_system}) by using the same trick which we used for $p=1$, namely by summing the $p+1$ relations (\ref{final_system}) corresponding to the different values of $r = 0, 1, \cdots, p$. One obtains
\beq\label{final_system_bis}
a_1\left(\frac{\sigma}{\sqrt{\lambda}} + c_p \frac{(p+1)(p+2)}{2} \right) + \sum_{\alpha=2}^{p+1} a_\alpha \frac{1}{t^*_\alpha}\left[\frac{1}{\left(1+t^*_\alpha\right)^{p+1}} + \frac{1}{\left(1-t^*_\alpha\right)^{p+1}}-2\right] =  c_p \frac{(p+1)(p+2)}{2} 
\eeq
where we have used the identity
\beq\label{identity_genp}
\sum_{r=0}^p \left[\frac{1}{(1-t)^{r+1}} - \frac{1}{(1+t)^{r+1}}\right] = \frac{1}{t} \left(\frac{1}{(1+t)^{p+1}} + \frac{1}{(1-t)^{p+1}} - 2  \right)
\eeq
which generalizes the identity (\ref{identity}) used for $p=1$. Since $t_\alpha^*$ satisfies the Eq. (\ref{eq:roots_eq}) with $s=1$ one has
\beq
(1-(t_\alpha^*)^2)^{p+1} - \frac{1}{2} \left[ (1+t_\alpha^*)^{p+1} + (1-t_\alpha^*)^{p+1} \right] = 0
\eeq
for $\alpha = 2,3, \cdots,p+1$ and therefore the coefficient of the term $\propto a_\alpha$ with $\alpha \geq 2$ in Eq. (\ref{final_system_bis}) vanishes, yielding an equation for the single unknown $a_1$. Finally, using the expression for $\sigma_p$ (\ref{var_p}) and $c_p$ (\ref{asympt_t1}), one has $c_p (p+1)(p+2)/2 = \sigma_p/\sqrt{2}$ and therefore, from Eq. (\ref{final_system_bis}), $a_1$ reads 
\beq\label{expr_a1_final}
a_1 = \frac{1}{1+ \sqrt{2/\lambda}} \;.
\eeq
Remarkably, this function $a_1$ is independently of $p$ -- the only $p$-dependence being contained in the scaling variable $\lambda = \sigma_p^2 (1-z)/\Delta^2$ through $\sigma_p$ (\ref{var_p}). From $a_1$, using Eq. (\ref{expansion_zto1_p}) and Eq. (\ref{stat_genp_GF}), one finally obtains that in the scaling limit $z \to 1$, $\Delta \to 0$ keeping $\lambda$ fixed, one has 
\beq\label{typ_GF_p}
\tilde p(z;\Delta) \sim \frac{\sigma_p^2}{\Delta^3} \frac{1}{(1+\sqrt{\lambda/2})^2} \;,
\eeq
as for $p=0$ and $p=1$ (\ref{typ_GF_p1}). Therefore, the gap PDF $p_k(\Delta)$ takes the same scaling form (\ref{scaling_form}), $p_k(\Delta) \sim \sqrt{k}/\sigma_p P(\sqrt{k} \Delta/\sigma_p)$ with the same density $P(x)$ given in Eq. (\ref{exact_F}), independently of $p$. This demonstrates the universality of this scaling function for this wide class of jump PDFs (\ref{f_eta_p}).

\subsection{Large fluctuations}

To study the large fluctuations of the limiting gap distribution $p_k(\Delta)$ in Eq. (\ref{stat_genp}) in the large $k$ limit, we study its GF $\tilde p(z;\Delta)$ in Eq. (\ref{stat_genp_GF}) in the limit $z \to 1$, but keeping $\Delta = O(1)$. This analysis can be carried out by generalizing the approach presented in Section \ref{sec:large_devp1} to any integer $p>1$. As explained there, we look for an expansion of $A_{\alpha,1}$ and $B_{\alpha,1}$ as given in Eq. (\ref{exp_large_dev_p1}) with $\alpha=0,1, \cdots, p+1$ which, once injected in the linear system (\ref{eq:sssto1}), yields a linear system of equations for the functions $A^{(0)}_{\alpha, 1}$, $A^{(1)}_{\alpha,1}$, $B^{(0)}_{\alpha, 1}$ and $B^{(1)}_{\alpha,1}$. The leading behavior of $\tilde p(z;\Delta)$ in the limit $z \to 1$ is then given by Eq. (\ref{large_p1}) with the substitution $c_1 \to c_p$ where $c_p$ is given in Eq.~(\ref{asympt_t1}). Finally, one finds that 
\beq\label{phi_p_genp}
p_{k}(\Delta) \sim \frac{1}{k^{3/2}} \varphi_p(\Delta) \;,
\eeq  
where $\varphi_p(\Delta)$ is given by Eq. (\ref{phi1_p1}), again with the substitution $c_1 \to c_p$. It is very hard to obtain an explicit expression for $\varphi_p(\Delta)$ but its asymptotic behaviors for $\Delta \to 0$ and $\Delta \to \infty$ can be obtained by analysing the aforementioned linear system for $A^{(0)}_{\alpha, 1}$, $A^{(1)}_{\alpha,1}$, $B^{(0)}_{\alpha, 1}$ and $B^{(1)}_{\alpha,1}$ in these two limits \cite{Matteo}. This yields
\begin{eqnarray}\label{asympt_phi1}
\begin{split}
\varphi_p(\Delta) \sim 
\begin{cases}
&3\sqrt{\frac{\left[\left(p+1\right)\left(p+2\right)\right]^3}{8\pi}}
\Delta^{-4} = \dfrac{3\,\sigma_p^3}{\sqrt{8 \pi}}\,\Delta^{-4} \;, \; \Delta\to0\\
& \\
&C_p\, \Delta^{2p}\,e^{-2\Delta} \;, \; \hspace*{1.85cm}\Delta \to\infty
\end{cases}
\end{split}
\end{eqnarray}
where $C_p$ is a constant, that we have not tried to compute for $p>1$ (it can be obtained by solving a rather complicated linear system).

\section{Conclusions}

In this paper, we have studied the PDF $P_{k,n}(\Delta)$ (\ref{def_pkn}) of the gaps $d_{k,n}$ (\ref{def_gap}) between the successive maxima of 
RWs of $n$ steps and whose jumps are distributed according to an Erlangen density with integer parameter $p+1$ (\ref{eq:2}). In the large $n$ limit, we have shown that $P_{k,n}(\Delta)$ converges to a limiting PDF $p_k(\Delta)$ that we have studied in detail in the limit of large $k$. Indeed, in this limit, $p_k(\Delta)$ exhibits two distinct regimes: (i) a typical regime for $\Delta =  O(1/\sqrt{k})$ and (ii) a large deviation regime for $\Delta = O(1)$. In the typical regime, we have obtained that $p_k(\Delta) \sim \sqrt{k}/\sigma_p P(\sqrt{k} \, \Delta/\sigma_p)$ where $\sigma^2_p$ is the variance of the jump PDF $f_p(\eta)$ (\ref{var_p}) and $P(x)$ is a universal and nontrivial scaling function given in Eq. (\ref{P_scaling}). In particular, for large $x$, it has an algebraic tail $P(x) \sim 3/\sqrt{8 \pi}\, x^{-4}$. We have demonstrated analytically that the scaling function $P(x)$ is independent of the jump PDF $f_p(\eta)$, thus confirming the conjecture about the universality of this function \cite{UOSRW} for a wide class of jump PDF (with finite variance). On the other hand, in the large deviation regime, we have shown that $p_k(\Delta) \sim k^{-3/2} \varphi_p(\Delta)$. Although the function $\varphi_p(\Delta)$ depends explicitly on $p$ (see for instance Eq. (\ref{asympt_phip})), it exhibits a singular behavior as $\Delta \to 0$,  $\varphi_p(\Delta) \propto \Delta^{-4}$ for all $p$, but with a $p$-dependent prefactor that ensures a smooth matching between the typical and the large fluctuations regime. 

These behaviors of $p_k(\Delta)$ for large $k$ have interesting consequences on the moments of the gaps, $\langle  d_{k,n}^m\rangle$, in the limit of large $n$ and large $k$, as it was already pointed out in \cite{UOSRW}. In fact, because of the algebraic tail of the scaling function $P(x) \propto x^{-4}$, we see that only the lowest moments, of order $m<3$, are dominated by the typical fluctuations which implies that $\langle  d_{k,n}^m\rangle \propto (\sigma_p/\sqrt{k})^m$, with a universal prefactor. On the other hand, for $m >3$ the moments are actually dominated by the large fluctuation regime of $p_k(\Delta) \propto k^ {-3/2}$, which implies that $\langle  d_{k,n}^m\rangle \propto k^{-3/2}$ with a non-universal amplitude. The moment of order $m=3$ receives contributions from both the typical and large deviation regime of $p_k(\Delta)$ and a careful analysis shows that it behaves as $\langle d_{k,n}^3\rangle \propto (\sigma_p/\sqrt{k})^3 \ln k$ with a universal prefactor \cite{Matteo}. These behaviors can be summarized as follows
\bea\label{moments}
\frac{\langle d_{k,n}^m\rangle}{\sigma_p^m} \sim
\begin{cases}
&\dfrac{1}{\sqrt{2 \pi}} k^{-1/2} \;, \; m = 1 \\
&\\
&\dfrac{1}{2} k^{-1} \;, \; m = 2 \\
&\\
& \dfrac{3}{4 \sqrt{2 \pi}}({\ln k})\,k^{-3/2} \;, \; m = 3 \\
&\\
& D_{m,p} \, k^{-3/2} \;, \; m > 3 \;,
\end{cases}
\eea 
where the amplitudes $D_{m,p}$ for $m>3$ depend explicitly on both $m$ and $p$. 

The results obtained here raise several questions left for future investigations. Here these results have been obtained
through a careful analysis of the integral equations in (\ref{eq:WH}), which can be solved explicitly only for a limited class of jump PDFs $f_p(\eta) \sim |\eta|^p e^{-|\eta|}$. The fact that the scaling function $P(x)$ is universal and holds for any jump PDF with a finite variance $\sigma^2$ suggests that there might exist a more general and elegant method to compute the scaling function $P(x)$ in Eq.~(\ref{P_scaling}) that does not require the explicit form of $f(\eta)$, as long as it has a finite variance. It would be nice if techniques from fluctuation theory for RWs could be used to obtain this universal scaling function \cite{Pitman}. Of course, it is natural to ask about the statistics of the gaps $d_{k,n}$ for jump PDF with a heavy tail, $f(\eta) \propto |\eta|^{-1-\mu}$ with a L\'evy index $\mu< 2$ (for which the variance $\sigma^2$ is infinite). Recently, exact results have been obtained for the first gap $d_{1,n}$ \cite{philippe, Philippe2,Philippe3} but obtaining results for higher order gaps $d_{k,n}$ with $k\geq 2$ seems quite challenging.

\ack

We would like to thank J. M. Luck, A. Dembo and J. Pitman for useful discussions. This research was supported by ANR grant ANR-17-CE30-0027-01 RaMaTraF.

\appendix

\section{Derivation of the formula given in Eq. (\ref{GF_pdf_delta})}\label{sec:app:64}

In this appendix, we derive the formula for the GF $\tilde P(z,s;\Delta)$ given in Eq. (\ref{GF_pdf_delta}) for $p=0$, which can then be generalized to any integer value of $p$ (\ref{eq:full_prob}). The starting point is the expression for $\tilde P(z,s;\Delta)$ given in Eq. (\ref{eq:6})
\begin{equation}\label{eq:6_app}
\begin{split}
\tilde{P}(z,s;\Delta) =& -\int_{0}^{\infty}dx\int_{x}^{\infty}dy\frac{\partial^2}{\partial x\partial y}\tilde Q(z,s;x,y-x)\delta(y-x-\Delta)\\ &- z\,\int_{-\infty}^{0}dx\int_{x}^{0}dy\frac{\partial^2}{\partial x\partial y}\tilde R(z,s;-y,y-x)\delta(y-x-\Delta) \;.
\end{split} 
\end{equation}
in terms of $\tilde Q(z,s;x,\Delta)$ and $\tilde R(z,s;x,\Delta)$ given respectively in Eq. (\ref{expl_qcal1}) and (\ref{expl_pcal1}). The formula (\ref{eq:6_app}) first requires the computation of $\tilde Q(z,s;x,y-x)$ and $\tilde R(z,s;-y,y-x)$ which read
\begin{eqnarray}\label{eq:6_1}
\begin{split}
&&\tilde Q(z,s;x,y-x) = \frac{1}{1-s} + A_1(y-x)\,e^{-t_1(s)\,x} \\
&&\tilde R(z,s;-y,y-x) = \frac{1}{1-zs} + B_1(y-x)\,e^{t_1(zs)\,y} \;.
\end{split}
\end{eqnarray}
where $t_1(s) = \sqrt{1-s}$. By differentiating these expressions with respect to both $x$ and $y$ we get
\begin{eqnarray}\label{eq:6_2}
\begin{split}
\frac{\partial^2}{\partial x \partial y} \tilde Q(z,s;x,y-x) &=& - (A_1''(y-x) +t_1(s) A'_1(y-x))e^{-t_1(s) \,x} \\
\frac{\partial^2}{\partial x \partial y} \tilde R(z,s;-y,y-x) &=& - (B_1''(y-x) +t_1(zs) B'_1(y-x))e^{t_1(zs) \,y}  \;.
\end{split}
\end{eqnarray}
Next, we insert these expressions (\ref{eq:6_2}) in Eq. (\ref{eq:6_app}) where the integrals over $x$ and $y$ can then easily been performed (in fact, because of the delta function, there is only a single integral to compute). It finally yields the expression given in Eq. (\ref{GF_pdf_delta}). This computation can be straightforwardly generalized to any $p$, by inserting in (\ref{eq:6_app}) the expressions for $\tilde Q$ and $\tilde R$ given in Eq. (\ref{eq:7}). In this way, we obtain the formula given in Eq. (\ref{eq:full_prob}).  

\section{Some useful integrals}\label{app:integrals}

The linear system of equations satisfied by the amplitudes $A_{\alpha}$ and $B_{\alpha}$ given in Eq. (\ref{eq:sss}) is obtained by inserting the expressions for $\tilde Q$ and $\tilde R$ given in Eq. (\ref{eq:7}) into the original integral equations (\ref{eq:WH}) satisfied by $\tilde Q$ and $\tilde R$. Hence we see that we need to evaluate the following integrals
\bea
&&{\cal J}_p^+(x,t) = \int_0^\infty e^{-t\,x'} f_p(x'+x) dx' \label{Jp} \\
&&{\cal J}_p^-(x,t) = \int_0^\infty e^{-t\,x'} f_p(x'-x) \label{Jm}\,dx' \;,
\eea  
where $f_p(\eta)$ is given in Eq. (\ref{eq:2}). To compute these integrals, it is convenient to compute their generating functions with respect to $p$. It is then straightforward, though a bit tedious, to obtain explicit expressions for ${\cal J}_p^+(x,t) $ and ${\cal J}_p^-(x,t) $~\cite{Matteo}. They read
\bea
&&{\cal J}_p^+(x,t) = \frac{e^{-x}}{2(t+1)^{p}} {I}_p(x(1+t)) \label{Jp1} \\
&&{\cal J}_p^-(x,t) = \frac{e^{-t\,x}}{2}\left(\frac{1}{(1-t)^p} +  \frac{1}{(1+t)^p}\right) - \frac{e^{-x}}{2(1-t)^{p+1}}I_{p}(x(1-t)) \;, \label{Jm1}
\eea
where we recall that $I_m(x) = \sum_{l=0}^m x^l/l!$. These explicit expressions (\ref{Jp1}) and (\ref{Jm1}) finally allow to derive the linear system of equations in (\ref{eq:sss}).


\end{document}